\documentclass[reprint,superscriptaddress,amsmath,amssymb,aps]{revtex4-1}
\usepackage[utf8]{inputenc}
\usepackage[T1]{fontenc} 
\usepackage{graphicx}
\usepackage{rotating}
\usepackage{makeidx}
\usepackage{threeparttable}
\usepackage{float}
\usepackage{amsmath}
\usepackage{rotating}
\usepackage{hyperref}
\usepackage{ulem}
\usepackage{relsize}
\usepackage{etoolbox}
\usepackage[most]{tcolorbox}
\usepackage{hyperref}
\usepackage{dcolumn}
\usepackage{bm}
\usepackage{wrapfig}
\usepackage{float}  
\usepackage[toc,page]{appendix}
\usepackage{gensymb}
\usepackage{epstopdf}
\usepackage{subfiles}
\usepackage{comment} 
\usepackage[autostyle=true]{csquotes} 
\usepackage{makecell}
\usepackage{siunitx}
\usepackage[all]{hypcap} 
\usepackage{booktabs}
\usepackage[most]{tcolorbox}
\usepackage{hyperref}
\usepackage{xcolor}
\usepackage{nicefrac}
\usepackage{chemformula}
\usepackage{ulem}
\definecolor{gainsboro}{rgb}{0.86, 0.86, 0.86}
\usepackage[backgroundcolor=gainsboro]{todonotes}
\hypersetup{
  colorlinks   = true,
  urlcolor     = blue,
  linkcolor    = blue,
  citecolor   = blue
}

\newcommand{\weg}[1]{}

\begin{document}
\title{Non-magnetic insulating phase induced by Jahn-Teller effect in RNiO$_3$}

\author{Sangeeta Rajpurohit} 
\email{rajpurohit1@llnl.gov}
\affiliation{Lawrence Livermore National Laboratory, Livermore, USA}
\affiliation{Molecular Foundry, Lawrence Berkeley National Laboratory, USA}

\author{Liang Z. Tan}
\affiliation{Molecular Foundry, Lawrence Berkeley National Laboratory, USA}

\author{Tadashi Ogitsu}
\affiliation{Lawrence Livermore National Laboratory, Livermore, USA}

\author{Peter E. Bl{\"o}chl} 
\affiliation{Institute for Theoretical Physics, Clausthal University of Technology, Germany}
\affiliation{Institute for Theoretical Physics, Georg-August-Universität Göttingen, Germany}

\date{\today}

\begin{abstract}
We propose a three-dimensional multi-orbital tight-binding model for rare-earth
nickelates RNiO$_3$ that treats charge, spin, orbital, and lattice degrees of freedom on equal
footing. All model parameters, including the on-site interactions $U$ and $J$ and the electron-phonon
(el-ph) coupling to the breathing mode, are extracted from hybrid-functional DFT calculations for
the small-bandwidth nickelate LuNiO$_3$. The model describes three competing insulating phases
governed by the interplay of $U{-}3J$ and el-ph coupling to the breathing and Jahn--Teller (JT) modes.
For large $U{-}3J$, the insulating state is stabilized by local JT distortions on high-spin
Ni$^{3+}$ sites. For smaller $U{-}3J$, the system undergoes charge disproportionation,
$2\mathrm{Ni}^{3+}\rightarrow\mathrm{Ni}^{2+}+\mathrm{Ni}^{4+}$, resulting in the spin-polarized
charge-ordered state observed experimentally below the N\'eel temperature in small-bandwidth RNiO$_3$.
When the JT energy on the Ni$^{2+}$ site exceeds Hund's exchange $3J$, a distinct charge- and
orbital-ordered insulating phase emerges in which the two $e_g$-electrons occupy the same orbital
with opposite spin. The stability of this phase is further confirmed by self-consistent calculations
within the full three-dimensional tight-binding model. This newly predicted metastable state,
characterized by JT distortions in a nonmagnetic charge-ordered RNiO$_3$ phase, shows that the
onset of magnetic order is not required for the metal-insulator transition in RNiO$_3$.
\end{abstract}

\maketitle

\section{Introduction}
Rare-earth nickelates RNiO$_3$ (where R is a rare-earth element)
belong to the class of strongly correlated transition metal
oxides with tightly coupled electronic, magnetic, and lattice
degrees of freedom. RNiO$_3$ undergoes a metal-insulator
transition (MIT) as the temperature decreases \cite{Medarde1997,Alonso1999, Alonso2001},
except large-bandwidth LaNiO$_3$, which remains metallic even at
low temperatures.  The MIT in these systems coincides with the
structural transition from $\rm{Pbnm}$ to $\rm{P2_1/n}$ symmetry.
The transition temperature $T_{MI}$ for larger-bandwidth RNiO$_3$
coincides with their magnetic transition temperature $T_{N}$ while
the magnetic transitions for smaller-bandwidth systems
occur at temperatures below their respective $T_{MI}$.
Given the experimental observations that indicate changes
in the electronic, magnetic, and structural arrangements
accompanying the MITs, the interpretation of the
underlying driving mechanism of the same remains
challenging from a theoretical perspective. 

Earlier theoretical studies of MIT in these systems focus on Mott-Hubbard or
p-d charge-transfer scenarios, depending on the relative
strength of the charge-transfer energy and the d-d Coulomb
interaction energy $U$ \cite{Alonso2001,Torrance1992,Imada1998,Stewart2011}.
Although Ni$^{3+}$ ions with nominal $e_g^1$ 
configurations are expected to display the Jahn-Teller (JT) effect; the
absence of the same in experimental studies is surprising. 
Mazin \textit{et al.}~\cite{Mazin2007} proposed that the relevant interaction
scale is not $U$ alone, but the reduced effective interaction $(U{-}3J)$, which becomes small in
the presence of strong Hund’s coupling. In this regime, Hund’s exchange competes with the local
Coulomb interaction and favors a charge-disproportionated insulating state, with the breathing
distortion providing an additional lattice stabilization. This interpretation was later supported
by studies identifying nickelates as small- or negative-charge-transfer systems 
\cite{Lau2013,Abbate2002,Han2012,Horiba2007,Mizokawa2000,Park2012}, where strong hybridization between
Ni-$e_g$ and O-$p$ states substantially screen the local Coulomb interaction and further reduce
the effective $(U{-}3J)$ scale. Within the negative charge-transfer framework, an alternative
ligand-hole description has been proposed in which charge is transferred from oxygen ligands to
neighboring Ni ions, yielding an effective disproportionation into $3d^8\underline{L}^2$ and
$3d^8$ configurations on the two inequivalent Ni sites \cite{Mizokawa2000_2,Varignon2017}. Here,
$\underline{L}$ denotes a ligand hole on the oxygen sublattice. The two Ni sites are distinguished
by their local octahedral environments, corresponding to compressed Ni$_S$ and expanded Ni$_L$
octahedra. This description is consistent with the $(1/2,1/2,1/2)$ superstructure peaks observed
in x-ray and neutron diffraction below the MIT, which indicate a rock-salt-like charge-ordered
state \cite{Alonso1999,Munoz1994,Scagnoli2005}. 
{While orbital degrees of freedom remain largely inactive in equilibrium, a few studies have shown that
epitaxial strain can modify the electronic and structural ground states of RNiO$_3$~\cite{Piel2014,He2015},
enhancing orbital polarization in LaNiO$_3$ and driving LuNiO$_3$ from a bond-disproportionated
charge-ordered state toward a Jahn--Teller-distorted metallic phase.

Previous theoretical studies suggest that the
insulating phase in bulk $\mathrm{RNiO_3}$ is intrinsically magnetic, implying that magnetic order
is required to stabilize the insulating state. This interpretation remains difficult to reconcile
with experiments on smaller-bandwidth RNiO$_3$, where the MIT occurs at temperatures $T_{MI}$
significantly above the onset of long-range magnetic order \cite{Medarde1997,Alonso1999,Alonso2001}.

Model Hamiltonians for rare-earth nickelates have been extensively studied to
describe the coupled structural and electronic instabilities associated with the MIT \cite{Zavaleta2021,Subedi2015,Lu2017,Ruppen2015,Lee2011}. 
Most of these works employ effective low-energy Hamiltonians that primarily emphasize electronic interactions, while lattice effects
are incorporated only indirectly through on-site potential differences between inequivalent Ni sites \cite{Subedi2015,Ruppen2015}.
The orthorhombic GdFeO$_3$-type tilts are similarly incorporated as a static on-site orbital field~\cite{Lee2011}. 
The possibility of JT distortions associated with orbital polarization remains largely unexplored within
such approaches. While these effective Hamiltonians capture the relevant low-energy physics, they do
not uniquely constrain the parameter space, as distinct parameter sets can reproduce similar observables.

In the present work, we propose a multi-orbital TB-model (TB) for RNiO$_3$.  The model parameters 
are derived directly from first-principles calculations. Within the atomic limit approximation of the TB-model
involving a pair of Ni$^{3+}$ ions and considering local effects such as Coulomb $U$, Hund's exchange $J$
and el-ph couplings, we elucidate the possible underlying MIT. In the absence of JT-type el-ph coupling,
the model predicts a $\rm{2Ni^{3+} \rightarrow Ni^{2+} + Ni^{4+}}$ charge-disproportionation 
for small $(U{-}3J)$, where the Ni$^{2+}$ site is in a high-spin $S{=}1$ state. The energy of this
charge-disproportionated configuration is further reduced by the distortion of the oxygen octahedra
surrounding the $\rm{Ni^{2+}}$ site. With increasing JT-type el-ph coupling, the energy gain
$\epsilon_{JT}$ of $\rm{e_g}$-electrons at the $\rm{Ni^{2+}}$ site due to JT distortion becomes comparable
to Hund's coupling, $J$. When $J/\epsilon_{JT}{<}2/3$, the model predicts an orbitally polarized
site $\rm{Ni^{2+}}$ more energetically favorable than the spin-polarized $\rm{Ni^{2+}}$.  

When we go beyond this atomic limit approximation to the three-dimensional TB-model,
both the charge disproportionation and the local orbital polarization at $\rm{Ni^{2+}}$
sites develop into a long-range charge and orbital ordering pattern. This
charge- and orbital-ordered (CO-OO) state is insulating with a finite band gap
at the Fermi level. The stabilization of the charge ordering due to the local JT-effect
in the absence of magnetism in the CO-OO state indicates that the onset of magnetism
is not necessary for MIT as suggested by previous theoretical studies
\cite{Alonso2001,Torrance1992,Imada1998,Stewart2011}. Our findings
are consistent with experimental studies that show that MIT temperatures $T_{MI}$
are higher than magnetic transition temperatures $T_{N}$ in smaller-bandwidth RNiO$_3$.

\section{Tight-Binding Model}
\label{secs:sec_2}
The electronic, structural, and magnetic properties of RNiO$_3$ perovskite
are investigated employing a 3d TB-model. Using a DFT-like framework, the total 
energy $E_{pot}$ of RNiO$_3$ system in our model is 
\begin{eqnarray}
E_{pot}\Big(|\psi_n\rangle,Q_{i,R}\Big)
&=&E_{el}(|\psi_n\rangle)+E_{ph}(Q_{i,R}) \nonumber\\
&&\hspace{-2cm}+
E_{el-ph}(|\psi_n\rangle,Q_{i,R}),
 \label{eq:tbm}
\end{eqnarray} 
which is a function of the electron wave functions $|\psi_n\rangle$ and lattice
distortions $Q_{i,R}$. It is the sum of the electron energy $E_{e}$, the energy due
to lattice distortions $E_{ph}$, and the el-ph coupling $E_{el-ph}$.

The electrons are described by a basis set of Ni-3d orbitals of $e_g$-character,
which are the orbitals pointing towards the neighboring oxygen ions. These
orbitals $|\chi_{\sigma,\alpha,\vec{R}}\rangle$ are characterized by a spin
index $\sigma\in\{\uparrow,\downarrow\}$, an orbital index $\alpha\in\{a,b\}$
and a site index $R$ specifying a specific Ni ion. The orbital $a$ is the
$x^2{-}y^2$ atomic orbital and the orbital b is the $3z^2{-}r^2$ atomic orbital
in the local coordinate system of the octahedron. While the basis orbitals are
characterized by the quantum numbers of a Ni-atom, each basis orbital has
an antibonding contribution from the oxygen orbitals built in. That is, the
oxygen contribution of the wave function is implicitly included by downfolding.
A wave function has the form
\begin{equation}
    |\psi_n\rangle=
     \sum_{R}
    \sum_{\sigma\in\{\uparrow,\downarrow\}}
    \sum_{\alpha\in\{a,b\}}
    |\chi_{\sigma,\alpha,R}\rangle 
    \psi_{\sigma,\alpha,R,n}
    \label{eqn:wavefunction}
\end{equation}
The phonon subsystem is described by two active JT modes, $Q_{2,R}$
and $Q_{3,R}$,  and one breathing mode $Q_{1,R}$ for an octahedron centered at
the Ni-site $R$. These modes are defined by the displacements of oxygen ions from
their equilibrium positions (see Appendix).

The kinetic energy $E_{hop}$ of $e_g$-electrons and the electron-electron (el-el) 
interaction $E_{coul}$ contribute to the energy $E_{el}=E_{hop}+E_{coul}$ of the
electronic system. The $e_g$-electrons delocalize between Ni sites through intermediate
oxygen bridges, and the kinetic energy term $E_{hop}$ is expressed as 
\begin{eqnarray}
E_{hop}=\sum\limits_{R,R',\sigma,n}
 f_n \sum\limits_{\alpha,\alpha'} \psi^*_{\sigma,\alpha,R,n}t_{\alpha,\alpha',R,R'}
  \psi_{\sigma,\alpha',R',n}
\label{eq:tbm_2}
\end{eqnarray} 
where $f_n$ are the occupations, $t_{\alpha,\alpha',R,R'}$ is the hopping matrix
element described in the appendix. The hopping matrix elements are limited to
neighboring Ni sites connected by an oxygen bridge.

The el-el interaction energy $E_{coul}$ is expressed as 
\begin{widetext}
\begin{eqnarray}
&E_{coul}&= \mathlarger{\mathlarger{\sum}}_{R}\Bigg(\frac{U}{2}\sum_{\substack{\sigma_1\ne\sigma_2 \\\alpha}}n_{\sigma_1\alpha,R}n_{\sigma_2\alpha,R}
{+}\rho_{\sigma_1\alpha,\sigma_2\alpha,R}\rho_{\sigma_2\alpha,\sigma_1\alpha,R}  
{+}\frac{U{-}3J}{2}\sum_{\substack{\sigma_1\\\alpha_1\ne\alpha_2}}n_{\sigma_1\alpha_1,R}n_{\sigma_1\alpha_2,R}
{-}\rho_{\sigma_1\alpha_1,\sigma_1\alpha_2,R}\rho_{\sigma_1\alpha_2,\sigma_1\alpha_1,R} \nonumber \\
&+&\frac{U{-}2J}{2}\sum_{\substack{\sigma_1\ne\sigma_2\\\alpha_1\ne\alpha_2}}n_{\sigma_1\alpha_1,R}n_{\sigma_2\alpha_2,R}
{-}\rho_{\sigma_1,\alpha_1,\sigma_2,\alpha_2,R}\rho_{\sigma_2,\alpha_2,\sigma_1,\alpha_1,R}  
{-}\frac{J}{2}\sum_{\substack{\sigma_1\ne\sigma_2\\\alpha_1\ne\alpha_2}}(\rho_{\sigma_1,\alpha_1,\sigma_2,\alpha_1,R}\rho_{\sigma_2,\alpha_2,\sigma_1,\alpha_2,R}
{-}\rho_{\sigma_1,\alpha_1,\sigma_1,\alpha_2,R}\rho_{\sigma_2,\alpha_2,\sigma_2,\alpha_1,R}) \nonumber \\
&{+}&\frac{J}{2}\sum_{\substack{\sigma_1\ne\sigma_2\\\alpha_1\ne\alpha_2}}(\rho_{\sigma_1,\alpha_1,\sigma_1,\alpha_2,R}\rho_{\sigma_2,\alpha_1,\sigma_2,\alpha_2,R}
{-}\rho_{\sigma_1,\alpha_1,\sigma_1,\alpha_2,R}\rho_{\sigma_2,\alpha_1,\sigma_2,\alpha_2,R})\Bigg)
\label{eq:tbm_2}
\end{eqnarray} 
\end{widetext}
where the on-site elements of the one-particle-reduced
density matrix used above is defined as 
\begin{equation}
\rho_{\sigma,\alpha,\sigma',\alpha',R}{=}\sum\limits_{n}
f_n\psi_{\sigma,\alpha,R,n}\psi^*_{\sigma',\alpha',R,n}.
\label{eq:densmat}
\end{equation}
The above expression of $E_{coul}$ contains terms similar to the local repulsion
terms in Kanamori Hamiltonian \cite{Junjiro1963,Jernej2013}. The first three terms,
with prefactors $U/2$, $(U{-}3J)/2$, and $(U{-}2J)/2$, correspond to electron-electron
interactions involving electrons with opposite spins in the same orbital, parallel spins
in different orbitals, and opposite spins in different orbitals, respectively. The last two
terms describe pair-hopping and spin-flip processes. The above expression for $E_{coul}$
is rotationally invariant in the eg-orbital and spin space. 

The el-ph coupling $E_{e-ph}$ is
\begin{equation}
E_{el-ph} = g_{JT} 
\sum_{R,\sigma}\sum_{\alpha,\beta} \rho_{\sigma,\alpha,\sigma,\beta,R}
M^Q_{\beta,\alpha}(Q_{1,R},Q_{2,R},Q_{3,R}). 
\label{eq:eq_el-ph} 
\end{equation}
Here $g_{JT}$ and $g_{br}$ are the el-ph coupling constants
and
\begin{eqnarray}
\mathbf{M}^Q(Q_{1,R},Q_{2,R},Q_{3,R})=
\left(\begin{array}{cc}
Q_{3,R} & Q_{2,R}\\Q_{2,R} & -Q_{3,R}
\end{array}\right)-{\bm 1}\frac{g_{br}}{g_{JT}}Q_{1,R}\;.
\end{eqnarray}
The energy of the phonon subsystem consists of restoring energy 
\begin{eqnarray}
E_{ph}&=&\frac{1}{2}k_{JT}\sum_{R}\Big(Q^2_{2,R}+Q^2_{3,R}+\frac{k_{br}}{k_{JT}}Q^2_{1,R}\Big).
\label{eq:eq_ph}
\end{eqnarray} 

Unlike most previous model studies, we take into account the breathing and JT modes on an equal
footing in our proposed model, allowing for a systematic investigation of the combined effect of
these distortions. We will study the present 3d-model with or without cooperative lattice distortion. 
In perovskite oxides, each oxygen ion is shared by two transition metal ions, so that the distortions
of two neighboring octahedra are strongly coupled through the shared oxygen ion. Hence, the octahedral
distortions, defined by local $Q_{i,R}$ modes in Eqn \ref{eq:eq_el-ph} and \ref{eq:eq_ph}, are highly
cooperative. In the absence of cooperative octahedral distortions, the $Q_{i,R}$ octahedral modes
around the Ni sites are considered independent variables.

\section{First-principles calculations of LuNiO$_3$}
\label{secs:sec_3}
To parameterize the above proposed TB-model for RNiO$_3$, we use density functional
theory (DFT) calculations for LuNiO$_3$, a representative small-bandwidth nickelate that exhibits
a robust antiferromagnetic (AFM) charge-disproportionated insulating ground-state. 

The DFT calculations for LuNiO$_3$ are performed using the
projector-augmented wave (PAW) method. A $1{\times}2{\times}1$
supercell of the space group P2$_1$ / n, containing 40 atoms
(eight formula units) and the experimental
lattice constants $a{=}5.12$ \AA, $b{=}5.51$ \AA, and $c{=}7.35$ \AA \cite{Alonso2001} is used. The
calculations used a $4{\times}2{\times}4$ k-point grid and a plane-wave cutoff energy of 40 Ry. The
calculations were carried out using the  hybrid functional PBE0r \cite{Sotoudeh2017,Eckhoff2020,Rajpurohit2024_3}. 
The PBE0r functional is a local hybrid functional derived from PBE0 \cite{Adamo1999}, in which a fraction of
the semilocal PBE exchange is replaced by exact exchange, i.e. 
\begin{equation}
E^{\mathrm{PBE0}}_{xc}=E^{\mathrm{PBE}}_{xc}+ a_x\left(E^{\mathrm{HF}}_{x}-E^{\mathrm{PBE}}_{x}\right),
\end{equation}
where $a_x$ sets the amount of exact exchange. This exchange correction in PBE0r is evaluated
in a local orbital basis and retained only for the on-site matrix elements of the correlated
subspace. The off-site exchange and the correlation energy remain at the PBE level \cite{Perdew1996}. In this sense,
PBE0r resembles DFT+$U$ in that the correction acts only on the on-site part of the correlated
subspace, while at the same time retaining the short-range exact exchange characteristic of screened
hybrid functionals. The PBE0r functional has been shown to give a reliable description of insulating ground states,
local moments, and bond disproportionation in correlated transition-metal oxides
\cite{Sotoudeh2017,Eckhoff2020,Rajpurohit2024_3}, and is therefore well suited for the present
study of rare-earth nickelates. The mixing factors for on-site exchange, taken from previous work
with PBE0r \cite{Sotoudeh2017}, are $a^{\rm{Lu}}_x{=}15\%$ for rare-earth Lu and $a^{\rm{O}}_x{=}10\%$
for oxygen. Rather than using a single value of $a^{\mathrm{Ni}}_x$, we examine
how the electronic structure of LuNiO$_3$ evolves as the on-site Fock-exchange admixture for Ni
atom is varied.

For $a^{\mathrm{Ni}}_x{>}1.5\%$, the relaxed structure exhibits two inequivalent
Ni sites with distinct average Ni-O bond lengths, consistent with a bond-disproportionated state,
accompanied by a breathing distortion of the NiO$_6$ octahedra. The relative stability of the magnetic
order is sensitive to the magnitude of  $a^{\mathrm{Ni}}_x$, as previously reported in  DFT+U
studies \cite{Hampel2017}.  For $a^{\mathrm{Ni}}_x {>}3.0\%$, the FM insulating state is lower in
energy than the AFM state, whereas for smaller $a^{\mathrm{Ni}}_x$ the AFM state becomes energetically
favorable. Table \ref{tab:dft_quantities} summarizes the band-gap, Ni-O bond lengths, and magnetic
moments at Ni sites for the AFM insulating state for different values of $a^{\mathrm{Ni}}_x$.
For $a^{\mathrm{Ni}}_x{>}5.0\%$, the average Ni--O bond lengths for the expanded Ni$_L$ and compressed
Ni$_S$ octahedra are in close agreement with the experimental values of 2.03~\AA\ and 1.94~\AA,
respectively \cite{Alonso2001}. The calculated local magnetic moments on the Ni$_L$ and Ni$_S$
sites are 1.35 (1.45) and 0.14 (0.13) $\mu_B$ for $a^{\mathrm{Ni}}_x{=}5.0\%$ (7.5\%). This strong
site disproportionation and the local magnetic moments are consistent with previous DFT+$U$
studies \cite{Varignon2017,Hampel2017,Prosandeev2012,Yamamoto2002}.

Figure~\ref{fig:fig_1} shows the density of states of the AFM state for three different values
of $a^{\mathrm{Ni}}_x$. The total density of states is projected onto Lu-$f$, Ni-$d$, and O-$p$ orbitals.
The lower valence band, extending from approximately $-6$ to $-4$ eV below the Fermi level, is dominated
by Lu-$f$ states. The O-$p$ bonding states, which exhibit substantial hybridization with Ni-$e_g$ orbitals,
 also contribute in this energy range. The Ni-$t_{2g}$ states are largely non-bonding and lie at higher
energy than the bonding O-$p$ states. The upper valence band near the Fermi level is predominantly of
Ni-$e_g$ character, with a smaller contribution from O-$p$ states. Increasing $a_x^{\rm Ni}$ shifts the
occupied Ni-$d$ states to lower energy and increases the band gap at the Fermi-level. 

\begin{table}[t]
\centering
\label{tab:ax_results}
\renewcommand{\arraystretch}{1.15}
\begin{tabular}{c c c c  c c}
\hline
$a_x^{\mathrm{Ni}}$ &
Band gap &
$\langle Ni_L{-}O\rangle$ &
$\langle Ni_S{-}O\rangle $ &
$m_{\mathrm{Ni_L}}$ &
$m_{\mathrm{Ni_S}}$ \\
(\%) & (eV) & (\AA) & (\AA) & ($\mu_B$) & ($\mu_B$) \\
\hline
2.5 & 0.46 & 2.019 & 1.956 & 1.22 & 0.14  \\
5.0 & 0.73 & 2.028 & 1.945 & 1.35 & 0.14  \\
7.5 & 0.97 & 2.035  & 1.939 & 1.45 & 0.13 \\
10 & 1.17 & 2.04  & 1.932 & 1.52 & 0.13  \\
\hline
\end{tabular}
\caption{DFT results for the relaxed AFM CO-SO state as a function of the on-site Fock-exchange admixture $a_x^{\mathrm{Ni}}$
on Ni atoms in the PBE0r functional.}
\label{tab:dft_quantities}
\end{table}

Figure~\ref{fig:fig_2} (top) shows the projected density of states resolved with Ni decomposed
into $t_{2g}$- and $e_g$-contributions at the Ni$_L$ and Ni$_S$ sites. The breathing distortion
lowers the majority-spin $e_g$-states at the Ni$_L$ site, which form the top of the valence band.
The bottom of the conduction band is formed by the empty $e_g$-states on the Ni$_S$ site, whereas
the higher conduction states involve the minority-spin $e_g$-states on the Ni$_L$ site.

\begin{figure}[!thp]
\begin{center}
\includegraphics[width=\linewidth]{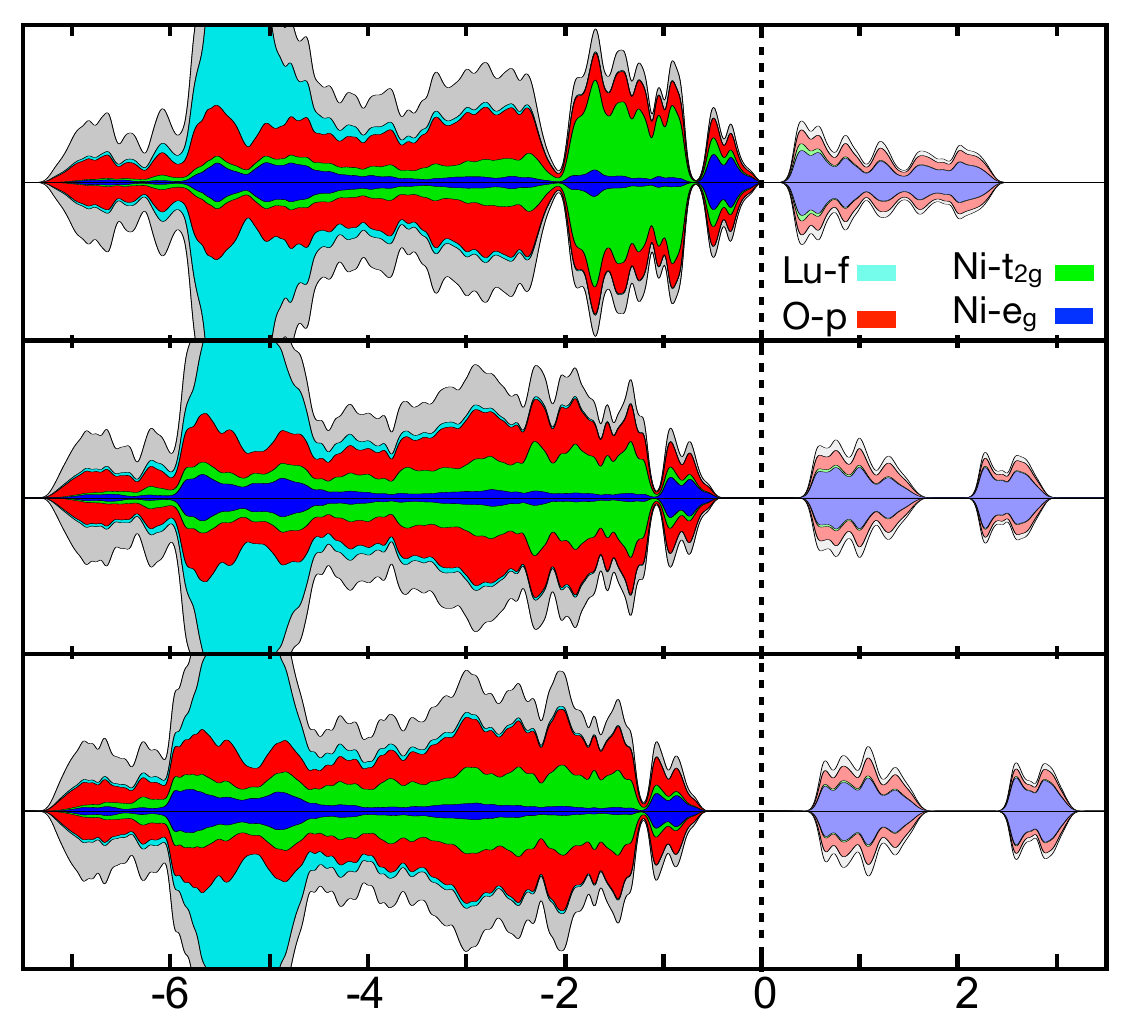}
\end{center}
\vspace{-0.5 cm}
\caption{
Total density of states (grey) of LuNiO$_3$ in the AFM ground-state, calculated
within DFT using the PBE0r hybrid functional, projected onto Ni-$d$ (green and blue),
Lu-$f$ (cyan), and O-$p$ (red) orbitals. The top, middle, and bottom panels correspond
to calculations performed for three values of the admixture parameter in PBE0r
$a_x^{\mathrm{Ni}}$: $2.5\%$ (top), $7.5\%$ (middle), and $10\%$ (bottom). The Ni-$d$
contribution is further resolved into Ni-$t_{2g}$ (green) and Ni-$e_g$ (blue) components.
The projected densities of states are stacked for clarity. The vertical dashed
line denotes the Fermi level for $a_x^{\mathrm{Ni}}=7.5\%$ case.}
\label{fig:fig_1}
\end{figure}

\section{Extraction of TB-model parameters}
To extract the parameters of the model introduced in the 
section \ref{secs:sec_2}, we optimize the phonon modes and compare the resulting 
electronic structure of the on-site TB-model with DFT results to constrain
$U$, $J$, $\varepsilon_{br}$, and $\varepsilon_{JT}$. Specifically, the energy
levels obtained from the optimized on-site Hamiltonian are matched to the first
moments (centers of mass) of the spin-resolved projected density of states
(PDOS) from the DFT ground-state. The hopping amplitude $t_{\mathrm{hop}}$ is
then fixed by requiring that the moments on the Ni$_L$ and Ni$_S$ sites obtained
from the full TB model reproduce the corresponding DFT calculated observables.

\subsection{On-site energy levels}
First, exploiting the isotropy of the on-site model in the
$(Q_{2,R},Q_{3,R})$ plane, we determine the optimal JT and breathing
distortions by minimizing the el-ph and elastic contributions
to the total energy, $E_{\mathrm{el-ph}}+E_{\mathrm{ph}}$, defined in
Eqs.~(\ref{eq:eq_el-ph}) and (\ref{eq:eq_ph}), with respect to the local
distortion modes, i.e.
$\partial(E_{\mathrm{el-ph}}+E_{\mathrm{ph}})/\partial Q_{i,R}=0$.
This gives
\begin{eqnarray}
Q_{3,R} &=& \frac{g_{JT}}{k_{JT}}\sum_\sigma(n_{\sigma b R}-n_{\sigma a R}), \\
Q_{1,R} &=& \frac{g_{br}}{k_{br}}\sum_{\sigma,\alpha}n_{\sigma\alpha R}.
\end{eqnarray}
The JT mode $Q_{3,R}$ is local to each NiO$_6$ octahedron, and
substituting the optimized distortion back gives the local JT energy
per site
\begin{equation}
E_{JT} = -\frac{\varepsilon_{JT}}{2}(\Delta n_R)^2,
\label{eq:C4}
\end{equation}
where $\Delta n_R=\sum_\sigma(n_{\sigma b R}-n_{\sigma a R})$ and
$\varepsilon_{JT}=g_{JT}^2/k_{JT}$.

\begin{figure}[!thp]
\begin{center}
\includegraphics[width=\linewidth]{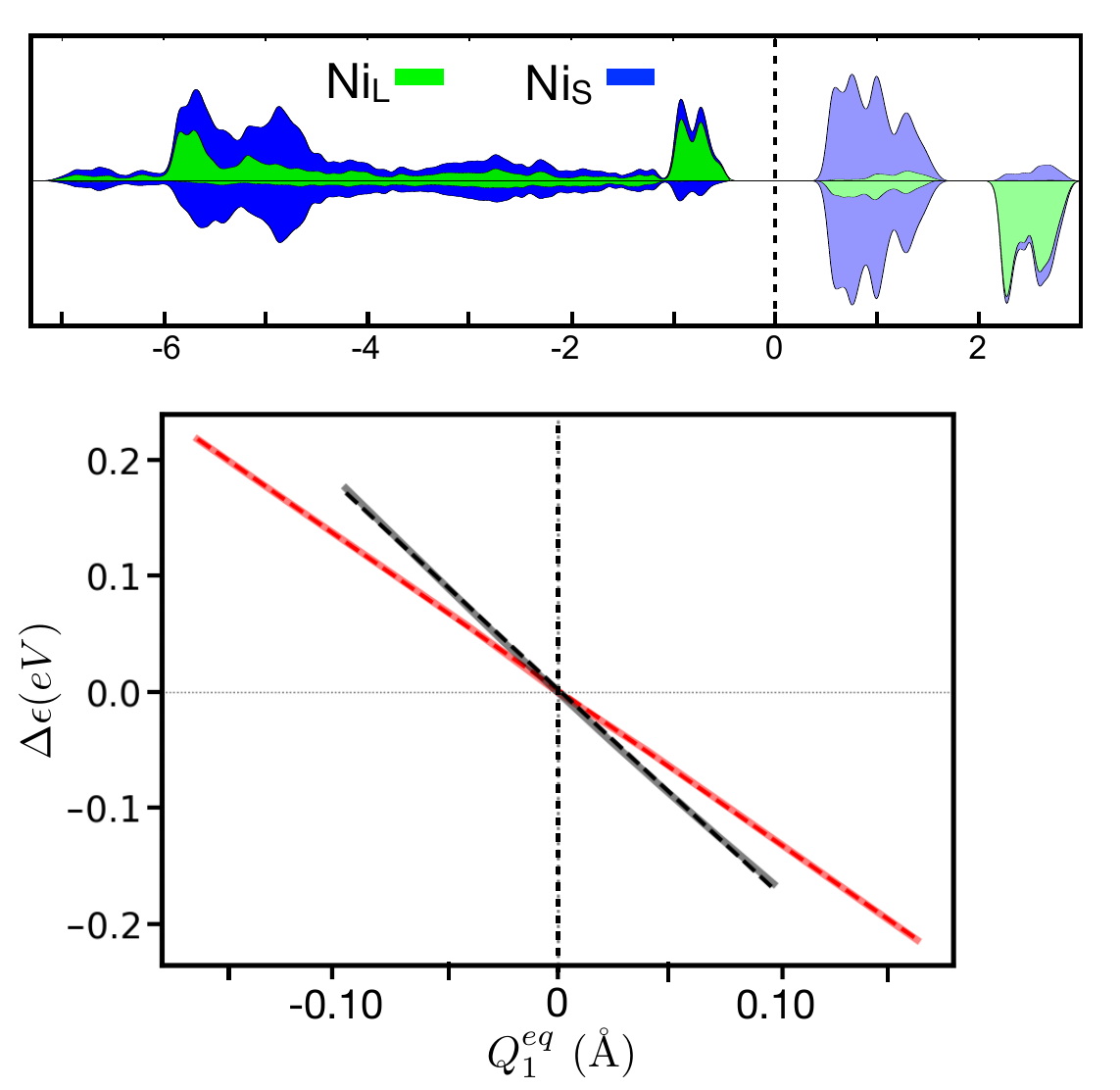}
\end{center}
\vspace{-0.5 cm}
\caption{Top: Density of states of LuNiO$_3$ in the AFM CO-SO state calculated
within DFT, projected onto $\mathrm{Ni_L}$-$e_g$ (blue) and $\mathrm{Ni_S}$-$e_g$
(green) states. The projected densities of states are stacked for clarity. The
vertical dashed line marks the Fermi level. Bottom: Shift $\varepsilon_{\uparrow b}^{L}-\varepsilon_{\sigma\alpha}^{S}$ 
 of the majority-spin $\mathrm{Ni_L}$-$e_g$ states w.r.t $\mathrm{Ni_S}$-$e_g$ states as a function
 of the breathing-mode distortion $\Delta Q_1^{eq}$ around the equilibrium structure for
 $a_x^{\mathrm{Ni}}$=5.0\% (red) and 7.5\% (black). The dashed lines show the corresponding
 linear fits with slopes 1.37 eV/\AA and 1.79 eV/\AA.}
\label{fig:fig_2}
\end{figure}

The breathing mode is cooperative: neighboring NiO$_6$ octahedra
share oxygen atoms, imposing $Q_{1}^{S}=-Q_{1}^{L}$ and reducing the
two-site problem to a single staggered coordinate $Q_1$ with corresponding energy
\begin{equation}
E_{br}(Q_{1})= -g_{br}Q_{1}(N_{L}-N_{S}) + k_{br}Q_{1}^2.
\label{eq:C5}
\end{equation}
Here $N_L-N_S=\sum_{\sigma,\alpha}(n_{\sigma\alpha L}-n_{\sigma\alpha S})$
is the difference in total $e_g$-occupation between the Ni$_L$ and
Ni$_S$ sites and both sites contribute to the elastic term.
Minimizing $E_{br}(Q_{1})$ over $Q_1$ gives $Q_1^{eq}=g_{br}(N_L{-}N_S)/2k_{br}$, and
substituting back in \ref{eq:C5} yields
\begin{equation}
E_{br}= -\frac{\varepsilon_{br}}{4}(N_L{-}N_S)^2,
\label{eq:C7}
\end{equation}
where $\varepsilon_{br}=g_{br}^2/k_{br}$.}

The fully optimized energy of the onsite model is then
\begin{align}
E[\rho] =\;
&\frac{U}{2}\!\sum_{\substack{R\\\sigma_1\ne\sigma_2\\\alpha}}
n_{\sigma_1\alpha R}n_{\sigma_2\alpha R}
+\frac{U-3J}{2}\!\sum_{\substack{R\\\sigma\\\alpha_1\ne\alpha_2}}
n_{\sigma\alpha_1 R}n_{\sigma\alpha_2 R} \nonumber\\
&+\frac{U-2J}{2}\!\sum_{\substack{R\\\sigma_1\ne\sigma_2\\\alpha_1\ne\alpha_2}}
n_{\sigma_1\alpha_1 R}n_{\sigma_2\alpha_2 R}
-\frac{\varepsilon_{br}}{4}(N_L{-}N_S)^2 \nonumber\\
&-\frac{\varepsilon_{JT}}{2}\sum_R(\Delta n_R)^2.
\label{eq:Erho}
\end{align}
We obtain the on-site electronic levels using Janak's theorem,
$\varepsilon_{\sigma\alpha R}=\partial E/\partial n_{\sigma\alpha R}$.
The cooperative breathing term in Eq.~(\ref{eq:C7}) contributes
$-\varepsilon_{br}(N_L-N_S)/2$ to all four on-site levels on the
Ni$_L$ site and the same term with opposite sign on the Ni$_S$ site.
The JT term in Eq.~(\ref{eq:C4}) contributes
$\mp\varepsilon_{JT}\Delta n_R$ to the $b$ and $a$ orbitals,
respectively. The four levels on the Ni$_L$ site are
\begin{align}
\begin{pmatrix}\varepsilon_{\uparrow b}\\ \varepsilon_{\uparrow a}\\
\varepsilon_{\downarrow b}\\ \varepsilon_{\downarrow a}
\end{pmatrix}^{L}
&=
-\frac{\varepsilon_{br}}{2}(N_{L}-N_{S})
\begin{pmatrix}1\\1\\1\\1\end{pmatrix}
-\varepsilon_{JT}\Delta n_L
\begin{pmatrix}1\\-1\\1\\-1\end{pmatrix} \nonumber\\
&+(U-3J)
\begin{pmatrix}N_{L}-n_{\uparrow b L}\\N_{L}-n_{\uparrow a L}\\
N_{L}-n_{\downarrow b L}\\N_{L}-n_{\downarrow a L}\end{pmatrix}
+J
\begin{pmatrix}n_{\downarrow a L}+n_{\downarrow b L}\\
n_{\downarrow a L}+n_{\downarrow b L}\\
n_{\uparrow a L}+n_{\uparrow b L}\\
n_{\uparrow a L}+n_{\uparrow b L}\end{pmatrix} \nonumber\\
&+2J
\begin{pmatrix}n_{\downarrow b L}\\n_{\downarrow a L}\\
n_{\uparrow b L}\\n_{\uparrow a L}\end{pmatrix},
\label{eq:onsite_levels}
\end{align}
with the sign of the breathing contribution reversed on the Ni$_S$ site.

Evaluating Eq.~(\ref{eq:onsite_levels}) in the CO-SO
ground-state ($N_L=2$, $N_S=0$, $S=1$, $\Delta n_L=0$), the Ni$_L$ levels are
orbitally degenerate within each spin channel,
\begin{align}
\varepsilon_{\uparrow b}^{L}=\varepsilon_{\uparrow a}^{L}
&= \varepsilon_0+(U-3J)-\varepsilon_{br}, \\
\varepsilon_{\downarrow b}^{L}=\varepsilon_{\downarrow a}^{L}
&= \varepsilon_0+(2U-2J)-\varepsilon_{br},
\end{align}
while all four Ni$_S$ levels remain degenerate,
$\varepsilon_{\sigma\alpha}^{S}=\varepsilon_0+\varepsilon_{br}$.
Two observable splittings follow directly: the spin splitting on
the Ni$_L$ site,
\begin{equation}
\varepsilon_{\downarrow b}^{L}-\varepsilon_{\uparrow b}^{L}=U+J,
\label{eq:split}
\end{equation}
and the shift of the Ni$_S$ level relative to the spin-averaged
Ni$_L$ center of mass,
$\bar{\varepsilon}^{L}=\frac{1}{2}(\varepsilon_{\uparrow
b}^{L}+\varepsilon_{\downarrow b}^{L})
=\varepsilon_0+\frac{3U-5J}{2}-\varepsilon_{br}$,
\begin{equation}
\varepsilon_{\sigma\alpha}^{S}-\bar{\varepsilon}^{L}=
2\varepsilon_{br}-\frac{3U}{2}+\frac{5J}{2}.
\label{eq:shift}
\end{equation}
Eqs.~(\ref{eq:split}) and (\ref{eq:shift}) provide two constraints
on $U$ and $J$ once $\varepsilon_{br}$ is fixed. Both are identified with the
first moments of the spin-resolved $e_g$ PDOS from the DFT ground-state. 
The values of $\varepsilon_{\uparrow b}^{L}$, $\varepsilon_{\downarrow b}^{L}$ and
$\varepsilon_{\sigma\alpha}^{S}$ obtained from DFT PDOS for different values
$a^{\mathrm{Ni}}_x$ and estimated $(U+J)$ are summarized in Table \ref{tab:mod_params}. 

\begin{table*}[t]
\centering
\renewcommand{\arraystretch}{1.15}
\begin{tabular}{c c c c c c c c c c}
\hline
$a_x^{\mathrm{Ni}}$ & $Q^{eq}_1$ &
$\varepsilon_{\downarrow b}^{L}-\varepsilon_{\uparrow b}^{L}$ &
 $\varepsilon_{\sigma\alpha}^{S}-\bar{\varepsilon}^{L}$ &
$g_{br}$ & $k_{br}$ & $(U+J)$ & $U$ & $J$ & $t_{hop}$ \\
 & \AA & (eV) & (eV) & (eV/\AA)  &  (eV/\AA$^2$) &  (eV)  & (eV) & (eV) & (eV) \\
\hline
5.0 \%& 0.280 & 2.075 & 0.087 & 1.37 & 9.79 & 2.075 & 1.37 & 0.70 & 0.25\\ 
7.5 \%& 0.332  & 2.547 & 0.047 & 1.79 & 10.78 & 2.547 & 1.73 & 0.82 & 0.275 \\ 
\hline
\end{tabular}
\caption{Energy differences and model parameters ($U$, $J$, $g_{br}$, $k_{br}$)
extracted from the Ni-$e_g$ projected density of states obtained from PBE0r calculations
at the equilibrium geometry, together with the resulting model parameters $U$, $J$,
and $\varepsilon_{br}$ as a function of the on-site Fock-exchange admixture
$a_x^{\mathrm{Ni}}$. Here, $Q_1^{eq}=\langle Q_1^L-Q_1^S\rangle_{eq}$ denotes
the equilibrium difference in breathing-mode between the Ni$_L$ and Ni$_S$ 
sites extracted from DFT, $\varepsilon_{\downarrow b}^{L}-\varepsilon_{\uparrow b}^{L}$
is the spin splitting at the Ni$_L$ site, and  $\varepsilon_{\sigma\alpha}^{S}-\bar{\varepsilon}^{L}$
is the Ni$_S$-Ni$_L$ level offset.}
\label{tab:mod_params}
\end{table*}

To extract the el-ph coupling $g_{br}$ to the breathing mode,
we perform DFT calculations to obtain the electronic structure at several
values of $Q_1^{L}$ spanning the equilibrium distortion
$Q_1^{eq}$. Figure~\ref{fig:fig_2} (bottom) shows the change in
$\varepsilon_{\uparrow b}^{L}-\varepsilon_{\sigma\alpha}^{S}$ as a
function of $\Delta Q_1^{eq}$. From the Janak
levels, this difference shifts linearly as
$-g_{br}\,\Delta Q_1^{eq}$, so the slope of a linear fit gives
 $g_{br}$. The spring constant $k_{br}$ then follows
from the cooperative equilibrium condition,
$k_{br}=2g_{br}/Q_1^{eq}$.
Substituting $\varepsilon_{br}=g_{br}^2/k_{br}$ into
Eqs.~(\ref{eq:split}) and (\ref{eq:shift}) yields
effective $U$ and $J$ for the model. 

The hopping amplitude $t_{hop}$ is chosen to reproduce
the DFT band gap within the three-dimensional TB-model.  In the fully relaxed
3d TB-model calculation, a slightly larger value of $g_{br}$ is required
to reproduce the DFT predicted $Q_1^{eq}$. In practice, we
increase $g_{br}$ from 1.79 to 2.15 eV/\AA, which accounts for the additional
charge-lattice feedback present in the self-consistent tight-binding calculation.
In Table \ref{tab:mod_params}, we summarize the extracted model parameters.

The extracted values of $U$ and $J$ are sensitive to
the choice of $a_x^{\rm Ni}$. The extracted $U{=}1.37$-$1.73$\,eV is
in good agreement with $U{\sim}1.85$\,eV from constrained random-phase
approximation (cRPA) calculations for a similar Ni-$e_g$
Hamiltonian~\cite{Hampel2019,Seth2017}, while the extracted
$J{=}0.70$-$0.82$\,eV is somewhat larger than the cRPA value
of $J{\sim}0.42$\,eV. In both cases, the extracted parameters
place LuNiO$_3$ well within the charge-disproportionated
insulating regime.

\subsection{Discussion of the parameter set}
Now we discuss three possible configurations for an average occupancy
of one $e_g$-electron per Ni site, depending on the relative magnitudes
of $U{-}3J$, $\varepsilon_{br}$, and $\varepsilon_{JT}$.

\textit{Spin- and orbital-polarized Ni$^{3+}$ (uniform JT state):}
For large $(U{-}3J)$, electrons remain equally distributed across Ni sites.
Each site develops both spin and orbital polarization and undergoes a local
JT distortion. The occupied $e_g$-orbital is lowered by $\varepsilon_{JT}$
through the el-ph coupling, while the elastic restoring energy $+\varepsilon_{JT}/2$
partially cancels this gain, giving a net energy of $-\varepsilon_{JT}/2$ per site.
This phase is stable when $\varepsilon_{br} < (U{-}3J) + \varepsilon_{JT}$.

\textit{Charge disproportionation with spin-polarized Ni$^{2+}$ (CO-SO state):}
For small $(U{-}3J)$, the system undergoes charge disproportionation,
$2\mathrm{Ni}^{3+}\rightarrow\mathrm{Ni}^{2+}+\mathrm{Ni}^{4+}$.
When $J/\varepsilon_{JT}{>}2/3$, the two $e_g$-electrons on the doubly
occupied Ni$^{2+}$ site align parallel ($S{=}1$) in accordance with
Hund's rule. In this charge- and spin-ordered (CO-SO) state, both
$e_g$-orbitals are equally occupied ($\Delta n_L{=}0$) and the JT
energy vanishes. The energy per Ni atom, measured relative to the
uniform metallic reference, is $E_{\mathrm{CO\text{-}SO}}/N_{\mathrm{Ni}}{=}(U{-}3J)/2-\varepsilon_{br}/2$,
where $(U{-}3J)/2$ is the Coulomb cost of placing two electrons on the
Ni$_L$ site and $\varepsilon_{br}/2$ is the breathing-mode energy gain per Ni,
which arises from the el-ph coupling $-g_{br}Q_1^{\mathrm{eq}}$ offset
by the elastic cost $+k_{br}(Q_1^{\mathrm{eq}})^2/2$. This phase is stable
when $\varepsilon_{br} > (U{-}3J) + \varepsilon_{JT}$.

\textit{Charge disproportionation with orbital-polarized Ni$^{2+}$ (CO-OO state):}
When the JT coupling dominates Hund's exchange, $J/\varepsilon_{JT}{<}2/3$,
the two electrons on the Ni$^{2+}$ site occupy the same orbital with
opposite spin ($S{=}0$). Both spins contribute to the orbital polarization,
giving $\Delta n_L{=}2$. The el-ph gain on the Ni$_L$ site is
$-4\varepsilon_{JT}$ while the elastic restoring energy is
$+2\varepsilon_{JT}$, giving a net JT energy of $-2\varepsilon_{JT}$
per Ni$_L$ site. The energy per Ni atom is
 $E_{\mathrm{CO\text{-}OO}}/N_{\mathrm{Ni}} = U/2 - \varepsilon_{br}/2-\varepsilon_{JT}$,
 where $U/2$ is the on-site Coulomb cost of the doubly-occupied $S{=}0$ configuration.

The boundary between the CO-SO and CO-OO phases is set by
$J/\varepsilon_{JT}{=}2/3$, while the boundary between the uniform JT and CO-SO
phases is given by $\varepsilon_{br} = (U{-}3J) + \varepsilon_{JT}$.
The boundary between the uniform JT and CO-OO phases is
$\varepsilon_{br}{=}U{-}\varepsilon_{JT}$. For the parameters extracted for
LuNiO$_3$, $(U{-}3J){<}0$, $\varepsilon_{br}{=}0.19$--$0.30$\,eV, and $\varepsilon_{JT}{<}3J/2$,
the system falls well inside the CO-SO regime, consistent with the
experimentally observed spin-polarized insulating ground state.
Since $(U{-}3J){<}0$, the condition $\varepsilon_{br}{>}(U{-}3J)$ is satisfied
for any positive $\varepsilon_{br}$, meaning that charge disproportionation
is driven by Hund's exchange rather than by the on-site Coulomb repulsion. The
condition $\varepsilon_{JT}{<}3J/2$ ensures that spin order is preferred over
orbital order at the Ni$^{2+}$ site.

The uniform JT and CO-SO phases, together with the competition between Hund's
coupling and on-site Coulomb repulsion as the driving force behind charge
disproportionation, were first discussed by Mazin et al.~\cite{Mazin2007}.
The nickelate phase diagram as a function of interaction strength and
el-ph coupling is studied in detail by Subedi
et al.~\cite{Subedi2015} and Peil et al.~\cite{Piel2019}, who showed
that proximity to an electronic disproportionation instability, coupled
to the breathing mode, controls the MIT across
the rare-earth series. The present model further allows for a CO-OO
phase, where charge disproportionation coexists with orbital, rather
than spin, order on the Ni$^{2+}$ site. This state was not identified
in the earlier studies and shows that spin polarization is not required
for a charge-disproportionated ground state in RNiO$_3$.

When hopping $t_{hop}$ is included, the kinetic energy gain from
electron delocalization differentially stabilizes the three phases.
In the uniform JT phase, all sites are equivalent, and hopping is unrestricted,
giving the largest band energy gain. In the CO-SO phase ($S{=}1$), the
large charge gap $\sim 2\varepsilon_{br}$ strongly suppresses inter-site
hopping. In the CO-OO phase ($S{=}0$), the two electrons occupy the same
orbital with opposite spins, leaving the orthogonal orbital available for
virtual hopping to the empty Ni$^{4+}$ site, providing an additional
kinetic energy gain absent in CO-SO. Hopping thus selectively stabilizes
CO-OO over CO-SO, effectively renormalizing the phase boundary from
$J/\varepsilon_{JT}{=}2/3$ to a smaller value, bringing the CO-OO phase
closer to realization in LuNiO$_3$.

\section{Full three-dimensional TB-model study}

The proposed on-site model neglects two important effects: the energy
gain by e$_g$-electrons due to delocalization between Ni
sites and the strain energy due to the cooperative nature of the
octahedral distortions. Previous studies \cite{GuzmanVerri2019} have
shown that the cooperative lattice effect significantly affects the phase diagram
in $\mathrm{RNiO_3}$.  In this section, we build upon the analysis of the
simple on-site model by extending it to a three-dimensional
TB-model. To investigate the combined effects of local and nonlocal
interactions, we compute the phase diagram of RNiO$_3$ using the full
3d TB model defined in Eq.~(1). We used a perovskite supercell with a
grid of $4\times4\times4$ Ni-ions and a grid of $4\times4\times4$ k-points. 
The parameters $t_{hop}$, $U$, $g_{br}$ and $k_{br}$ are
provided in Table~\ref{tab:mod_params}. Furthermore, we use
$g_{JT}{=}g_{br}$ and $k_{JT}{=}k_{br}$. For the rest of the paper,
we use the parameters extracted from the DFT calculations with
$a_x^{\mathrm{Ni}}=7.5\%$.

\begin{table}[t]
\centering
\renewcommand{\arraystretch}{1.15}
\begin{tabular}{l S[table-format=1.2] S[table-format=-1.2]}
\toprule
 & \multicolumn{2}{c}{$\Delta E = E_{tot} - E_{\mathrm{CO\text{-}SO}}$ (eV)} \\
\cmidrule(lr){2-3}
State      & {$g_{br}{=}3.225$} & {$g_{br}{=}3.76$} \\
\midrule
CO-OO       & 0.28 & -0.02 \\
Uniform JT  & 0.35 &  0.30 \\
\bottomrule
\end{tabular}
\caption{Energy difference $\Delta E = E_{tot} - E_{\mathrm{CO\text{-}SO}}$ (in eV)
for the CO-OO and uniform Jahn–Teller states, relative to the CO-SO reference state, as a function
of the breathing-mode coupling $g_{br}$ (in eV/\AA). Each state is fully relaxed, both
electronically and structurally.}
\label{tab:energy_phases}
\end{table}

To determine the ground-state and low-lying metastable
states, the total energy functional of the model is minimized with respect
to all dynamical degrees of freedom. This minimization is performed using
Car-Parrinello molecular dynamics as an optimization scheme for obtaining
minimum-energy configurations. The energy functional defined in Eq.~\ref{eq:tbm}
is minimized by evolving the electronic wavefunctions $|\psi_n\rangle$ and
lattice degrees of freedom $Q_{i,R}$ according to damped equations of motion.
A friction term for damping is included to dissipate energy and drive the
system toward a local minimum.

\begin{figure}[!thp]
\begin{center}
\includegraphics[width=0.75\linewidth]{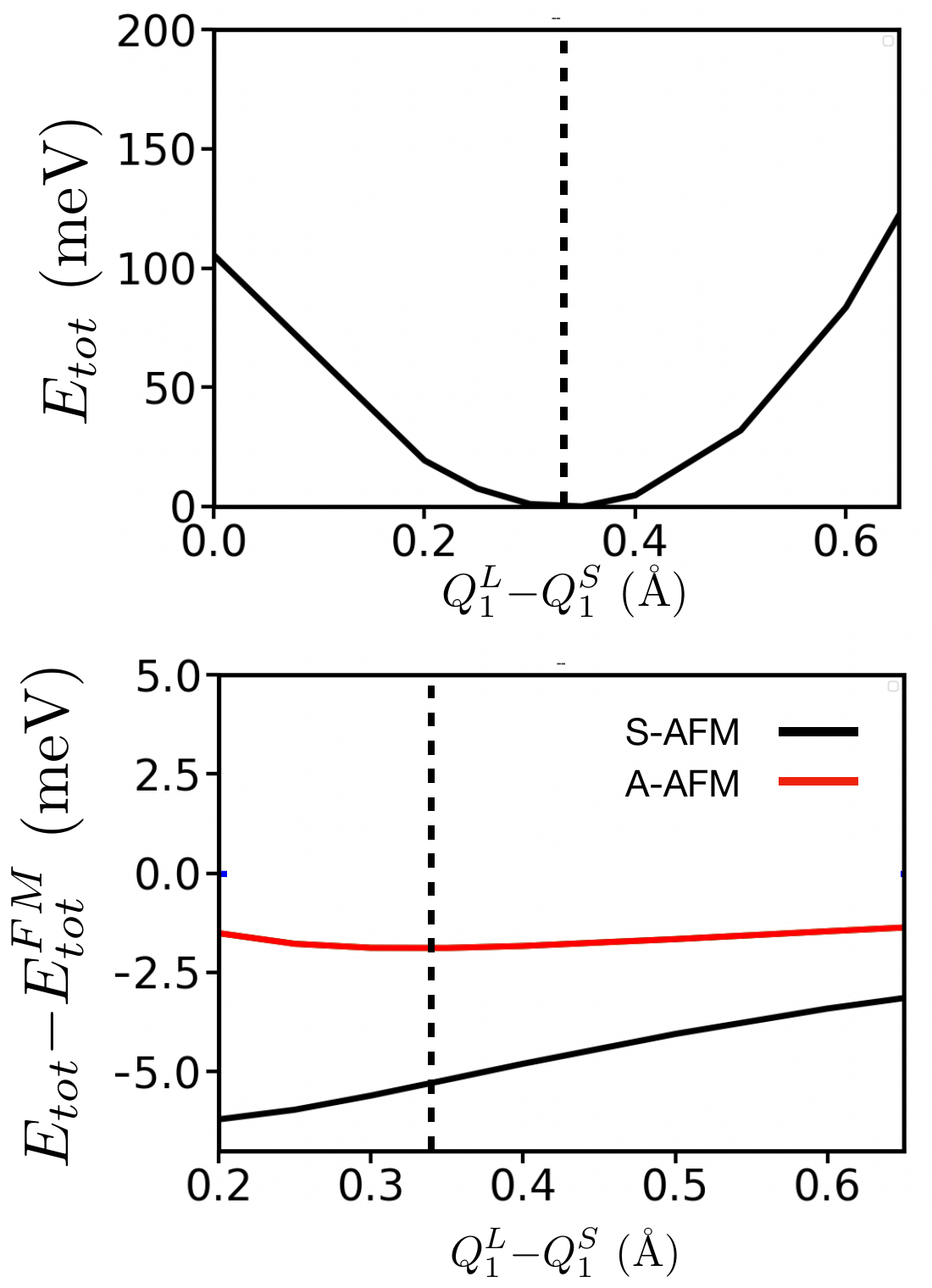}
\end{center}
\vspace{-0.5 cm}
\caption{ Top: TB-model predicted total energy as a function of the
breathing-mode $(Q_1^L-Q_1^S)$ in the CO-SO phase, referenced to the equilibrium
geometry. Bottom: TB-model total energy of the S- and A-type AFM spin configurations
in the CO-SO phase, referenced to the FM state, as a function of $(Q_1^L-Q_1^S)$ .}
\label{fig:fig_3}
\end{figure}

We calculate the total energy $E_{\mathrm{pot}}[\psi,Q_{i,R}]$ of RNiO$_3$
in the AFM and FM state within the 3d tight-binding model by self-consistently
relaxing the electronic structure. During initial optimization, S-type and
FM spin-order \cite{Giovannetti2009} is imposed by applying spin-dependent
external potentials acting on electrons at Ni sites. In the S-type
order, the Ni spins within the xy-plane form ferromagnetic zigzag chains.
The neighboring zigzag chains in the xy-planes are antiferromagnetically
coupled. In the c-direction, alternating ferromagnetic and antiferromagnetic
coupling between xy-planes results in a doubled periodicity. These potentials are
removed close to convergence. Figure~\ref{fig:fig_3} (top) shows the total
energy of the CO-SO state as a function of the breathing-mode distortion.
Figure~\ref{fig:fig_3} (bottom) shows the energetics of different spin configurations
in the CO-SO state as a function of the breathing-mode distortion. The AFM state
is clearly stabilized in the vicinity of the equilibrium value $Q_1$.

\begin{figure}[!thp]
\begin{center}
\includegraphics[width=0.75\linewidth]{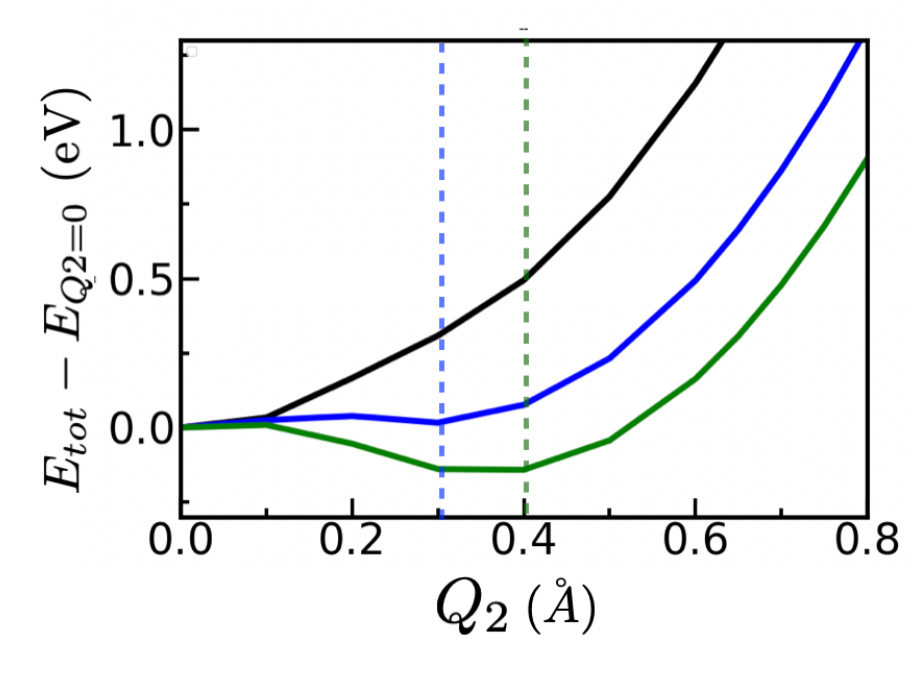}
\end{center}
\vspace{-0.5 cm}
\caption{TB-model predicted total energy of the uniform JT phase as a
function of the JT-mode $Q_2$ shown for different values of $g_{br}$. The black, blue, and
green curves correspond to $g_{br}{=}2.15$, $3.225$, and $3.76$ eV/\AA, respectively. The
vertical dashed lines indicate the energy minima of the uniform JT phase.}
\label{fig:fig_4}
\end{figure}

Although neither the uniform JT phase nor the CO-OO phase is stable
at the original value of $g_{br}$, increasing $g_{br}$ by a factor of 1.5 makes both phases
metastable. The energy differences of the fully relaxed CO-OO and uniform JT states relative
to the CO-SO state are listed in Table~\ref{tab:energy_phases}. The calculated values of
$Q_2$ for the uniform JT phase are $0.30$ \AA and $0.40$ \AA\ (see Figure \ref{fig:fig_4}),
obtained for $g_{br}$ enhanced by factors of 1.5 and 1.75, respectively.

\subsection{Charge- and spin-ordered state (CO-SO)}
The TB-model predicted AFM ground state exhibits checkerboard-type
charge order in all three directions, with two inequivalent Ni sites:
Ni$^{2+}$ (Ni$_L$) sites with large magnetic moments and nonmagnetic
Ni$^{4+}$ (Ni$_S$) sites. The local $e_g$ occupancy at the Ni$_L$
and Ni$_S$ sites is 1.83 and 0.17 $e$, respectively, and the
predicted breathing-mode distortion $Q_1^L{-}Q_1^S{=}0.39$\,\AA\
is in close agreement with the DFT value.

\begin{figure}[!thp]
\begin{center}
\includegraphics[width=\linewidth]{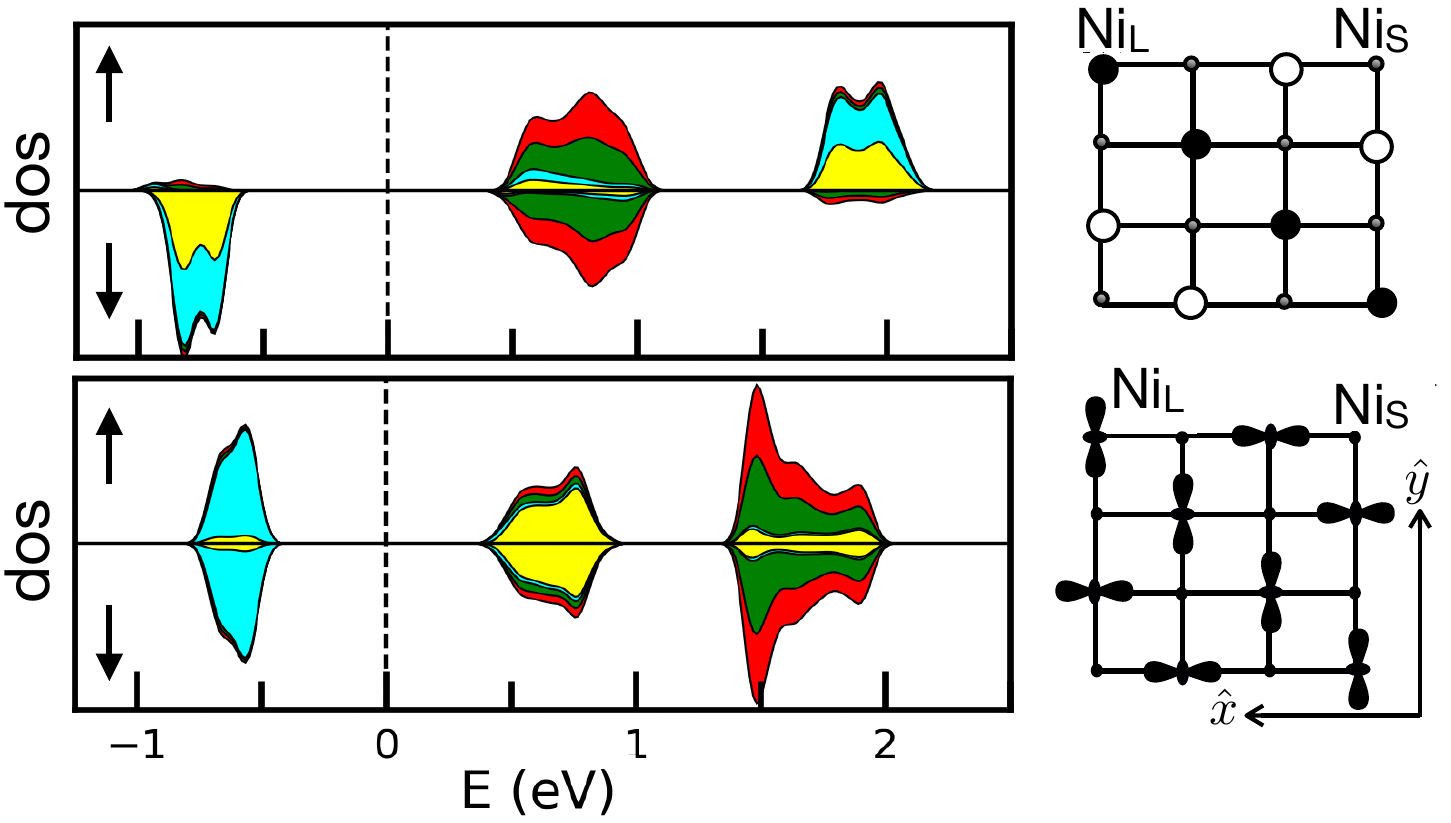}
\end{center}
\vspace{-0.5 cm}
\caption{
Left: Density of states of the CO-SO (top) and CO-OO (bottom)
structures predicted by the 3D TB model using
$g_{br}{=}2.15$~eV/\AA\ and $g_{br}{=}3.225$~eV/\AA,
respectively. The densities of states are projected onto the
Ni-$e_g$ orbitals of the Ni$_L$ (cyan and yellow) and
Ni$_S$ (red and green) sites. The projected densities of
states are stacked for clarity. The vertical dashed line marks
the Fermi level. In the CO-OO state, the orbital shown in cyan
is oriented along the longer Ni--O bond in the $xy$ plane,
while the orthogonal orbital is shown in yellow.
Right: Corresponding long-range charge, spin, and orbital order
in the $xy$ plane. In the CO-SO state, the large black and
white circles denote spin-up and spin-down Ni$_L$ sites,
respectively, while the smaller circles represent nonmagnetic
Ni$_S$ sites.}
\label{fig:fig_5}
\end{figure}

The density of the states of the predicted CO-SO phase of the model projected on
the Ni$^{2+}$ and Ni$^{4+}$ orbitals is shown in Figure~\ref{fig:fig_5} (top). The occupied
bands below the Fermi level predominantly consist of two eg-orbitals of
Ni$^{2+}$ sites and are spin-polarized. The conduction band consists
of Ni$^{4+}$ states, and there is no spin splitting at these sites. 
Above Ni$^{4+}$ states are the opposite spin $e_g$-states of Ni$^{2+}$ sites.

\subsection{Charge- and orbital-ordered state (CO-OO)}
To explore the stability of the CO-OO phase predicted by the on-site model, we examine
its energetics within the full 3d TB-model. Starting from the CO-OO state, we impose
the orbital pattern shown in Fig.~\ref{fig:fig_5} (bottom) and relax the electronic and
atomic structure. We vary the el-ph couplings by scaling both $g_{br}$ and $g_{JT}$ relative
to their extracted values and compare the energies of the CO-OO, CO-SO, and uniform
JT-distorted states.

\begin{figure}[!thp]
\begin{center}
\includegraphics[width=0.75\linewidth]{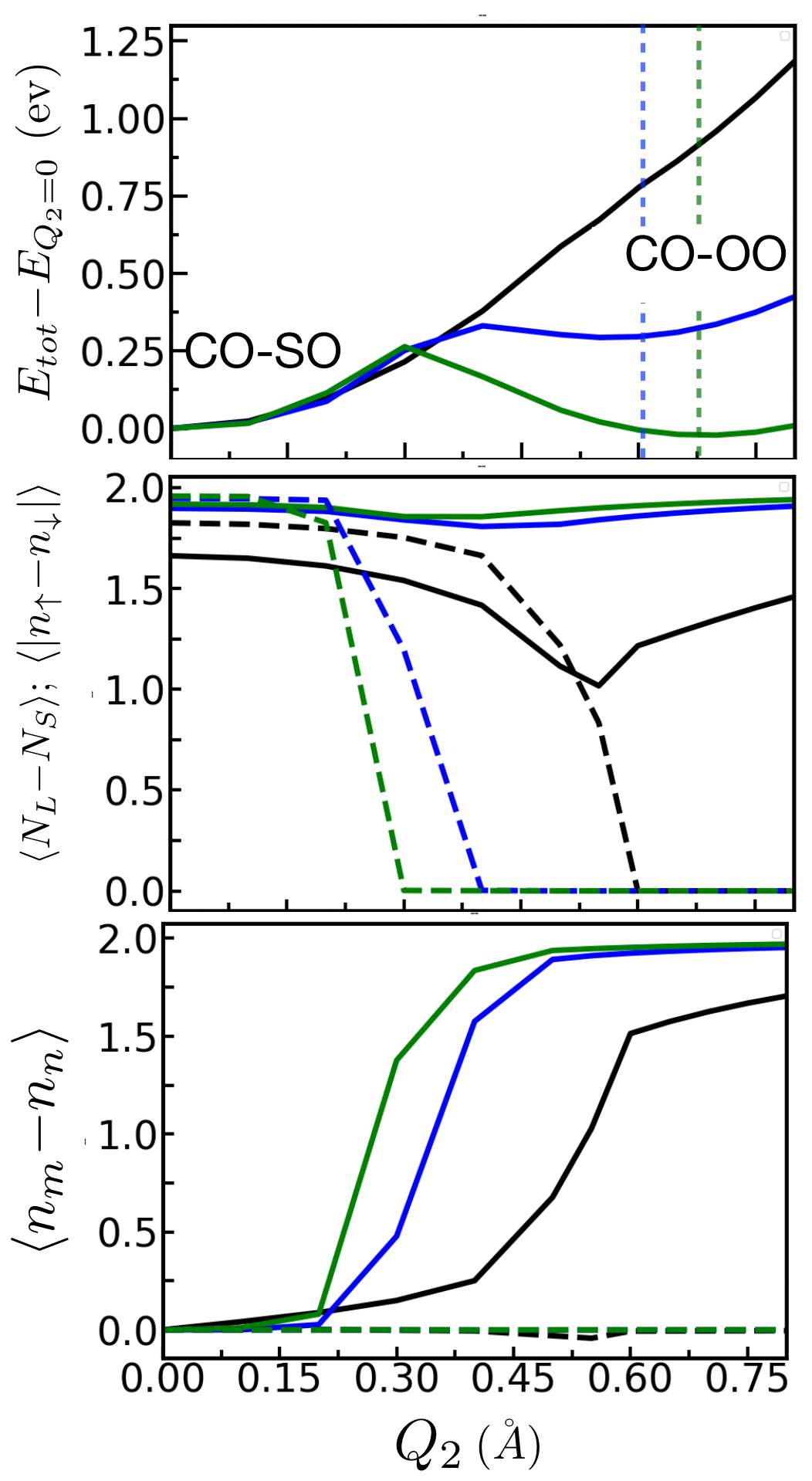}
\end{center}
\vspace{-0.5 cm}
\caption{TB-model calculated total energy as a function of the JT mode
$Q_2$ (top); Site difference between the Ni$_L$ and Ni$_S$ sites (middle), sites with charge
disproportionation difference shown in solid lines and magnetic-moment difference shown in dashed
lines; Orbital polarization $\langle n_m{-}n_n\rangle$ on Ni$_L$ and Ni$_S$ (bottom), shown as solid and
dashed lines, respectively. Results are shown for different values of $g_{br}$. The black, blue, and
green curves correspond to $g_{br}{=}2.15$, $3.225$, and $3.76$~eV/\AA, respectively. The vertical
dashed lines in the top panel indicate the energy minima of the CO-OO phase.}
\label{fig:fig_6}
\end{figure}

Figure~\ref{fig:fig_6} (top) shows the total energy as a function of the
JT-mode distortion at the Ni$_L$ site. For the parameters extracted for LuNiO$_3$, the
CO-SO phase is the lowest-energy solution. Upon increasing both $g_{br}$ and $g_{JT}$ by
approximately 50\%, the CO-OO phase develops as a metastable local minimum. In this regime,
its energy lies below that of the uniformly JT-distorted state, while remaining above the
CO-SO ground state. When $g_{br}$ is increased to  1.75 times its DFT-extracted value, the CO-OO
state with finite $Q_2$ becomes energetically favorable relative to the CO-SO state at
$Q_2{=}0$. The $Q_2$ distortion of the CO-OO state is significantly larger than the
corresponding values reported in strained RNiO$_3$ \cite{He2015}, but remains comparable in
magnitude to the JT mode found in other transition-metal oxides, such as PrMnO$_3$ with the same $e_g^1$
configuration~\cite{Sotoudeh2017}. As shown in Figure~\ref{fig:fig_6} (middle), the charge
disproportionation $(N_L{-}N_S)$ between the Ni$_L$ and Ni$_S$ sites remains nearly unchanged
relative to its value in the CO-SO phase. By contrast, the magnetic moment on the Ni$_L$ sites
collapses in the CO-OO phase, consistent with its nonmagnetic character. Simultaneously, a
finite orbital polarization develops, as shown in Figure~\ref{fig:fig_6} (bottom).

In the CO-OO state, the Ni$_L$ sites are nonmagnetic, accompanied by the simultaneous 
emergence of orbital polarization.  Figure~\ref{fig:fig_6} (bottom) shows a nonzero
orbital polarization $\langle n_m - n_n \rangle$ at the Ni$_L$ sites in the CO-OO state,
where $n_{m,R}$ and $n_{n,R}$ denote the occupancies of two
orthonormal $e_g$ orbitals $|m\rangle$ and $|n\rangle$ at site $R$.
The orbital $|m\rangle$ has lobes pointing along the direction of the
elongated Ni$_L$--O bond in the $xy$ plane, induced by the local $Q_2$
JT distortion, while $|n\rangle$ is the orthogonal $e_g$ orbital.
In the CO-OO state, $|m\rangle$ is predominantly occupied while
$|n\rangle$ remains largely empty on the Ni$_L$ site, giving rise to
a finite $\langle n_m-n_n \rangle > 0$.

Figure \ref{fig:fig_5} (bottom) left shows the density of the states of this CO-OO state projected on
the $e_g$-orbitals of Ni$_L$ and Ni$_S$ sites. Clearly, spin-splitting is absent at the Ni$_S$ and
Ni$_L$ sites. The JT-effect splits the orbitals at Ni$_L$ sites where the lobes of the lower occupied orbital,
indicated as Ni$_L$, point towards the larger Ni$_L$-O bonds. A finite band-gap at the Fermi level
suggests the insulating nature of the CO-OO state.

Our model predicts the CO-OO state to become stable only when
el-ph coupling is increased to approximately $1.75$ times the value extracted from DFT for
LuNiO$_3$. Such an enhancement of the el-ph coupling can arise from an increase in $g_{br}$, 
which varies across the rare-earth nickelate series RNiO$_3$, depending on the choice of
the rare-earth element, as suggested by Peil \textit{et al.}~\cite{Piel2019}. Epitaxial strain
provides an additional route to tune the el-ph coupling~\cite{Piel2014,He2015}. Alternatively,
ultrashort laser pulses can transiently reduce Hund's exchange $J$, thereby lowering the
effective $(U{-}3J)$, as suggested in recent theoretical and experimental
studies~\cite{Rajpurohit2026,Baykusheva2022,Dejean2018,Graanas2022}, and potentially
bringing the system closer to the CO-OO phase boundary.

Most theoretical studies in the past have suggested that
the onset of a magnetic order is necessary to stabilize the
insulating phase in RNiO$_3$ \cite{Varignon2017, Mazin2007}. This
is in contrast to experimental studies \cite{Medarde1997,Alonso1999, Alonso2001}
that show that the transition temperature $T_{MI}$ from the paramagnetic
metallic phase to the insulating CO-state is higher than the temperature
$T_{N}$ for smaller-bandwidth systems, suggesting that magnetism is not
the primary mechanism of MIT. The CO-OO state predicted from the
TB-model, resembles the experimentally observed paramagnetic insulating state: both
do not have any net magnetism and exhibit CO with $P2_1/n$ symmetry.
The previous experimental studies do not show the signature of any
long-range OO in a high-temperature paramagnetic insulating phase. This means
that the long-range order of the orbital-polarization predicted in the CO-OO
state must vanish, forming an "orbital-liquid" phase at temperatures
$T_{N}{<}T{<}T_{MI}$. It would be interesting to study in future work
whether the finite-temperature lattice dynamics destroys the
non-local orbital correlations of the proposed CO-OO state.  

\section{Summary}
We propose a three-dimensional tight-binding model
for RNiO$_3$ and extract its parameters from ab-initio calculations for a
smaller bandwidth LuNiO$_3$.  The TB-model treats all relevant degrees
of freedom, namely charge, spin, orbital, and lattice, explicitly. Unlike previous
models, we consider both octahedral breathing and JT modes in our model  to
investigate the combined effect of these modes in MIT in RNiO$_3$.
The extracted $U{=}1.37$--$1.73$\,eV is in good agreement with
$U{\sim}1.85$\,eV derived from constrained random-phase approximation
(cRPA) for a similar $e_g$-orbital low-energy model of RNiO$_3$~\cite{Hampel2019,Seth2017}, 
while the extracted $J{=}0.70$--$0.82$\,eV is approximately twice the cRPA value of
$J{\sim}0.42$\,eV.  For LuNiO$_3$, the extracted parameters, $(U{-}3J){<}0$ and
$\varepsilon_{JT}{<}3J/2$, place the system well inside the charge-ordered
spin-polarized (CO-SO) regime, consistent with the experimentally observed insulating ground state.

We further explore the phase space of the model by varying the el-ph couplings $g_{br}$ and $g_{JT}$
associated with the breathing and JT modes, respectively. Increasing these couplings by a factor
of 1.5-1.75 leads to the emergence of a new metastable insulating state with charge disproportionation
between Ni sites but no magnetic order, which we denote as CO-OO. The model-predicted CO-OO state
exhibits strong $e_g$-orbital polarization on the Ni$^{2+}$ sites. This local $e_g$-orbital polarization
within the charge-ordered state gives rise to a long-range orbital ordering pattern. The stabilization
of an insulating charge-ordered state by the local JT effect in the absence of magnetism shows that
the onset of magnetic order is not required for the metal-insulator transition,
contrary to previous theoretical studies.

\section*{Acknowledgments}
Theory and simulation were supported by the Computational Materials Sciences (CMS) Program
funded by the US Department of Energy, Office of Science, Basic Energy Sciences, Materials
Sciences and Engineering Division. Data analysis was provided by the User Program of the
Molecular Foundry, supported by the Office of Science, Office of Basic Energy Sciences, of
the U.S. Department of Energy under Contract No. DE-AC02-05CH11231. This work was performed
under the auspices of the U.S. Department of Energy by Lawrence Livermore National Laboratory
under Contract DE-AC52-07NA27344. S.R and T.O. are supported by the Computational Materials
Sciences Program funded by the US Department of Energy, Office of Science, Basic Energy Sciences,
Materials Sciences and Engineering Division. This work was partially funded by the Deutsche
Forschungsgemeinschaft (DFG, German Research Foundation)- 217133147/SFB 1073, project B03 and C03.
\bibliography{ref}

@Article{Sotoudeh2017,
  title = {Electronic structure of ${\mathrm{Pr}}_{1\ensuremath{-}x}{\mathrm{Ca}}_{x}{\mathrm{MnO}}_{3}$},
  author = {Sotoudeh, Mohsen and Rajpurohit, Sangeeta and Bl\"ochl, Peter and Mierwaldt, Daniel and Norpoth, Jonas and Roddatis, Vladimir and Mildner, Stephanie and Kressdorf, Birte and Ifland, Benedikt and Jooss, Christian},
  journal = {Phys. Rev. B},
  volume = {95},
  issue = {23},
  pages = {235150},
  numpages = {23},
  year = {2017},
  month = {Jun},
  publisher = {American Physical Society},
  doi = {10.1103/PhysRevB.95.235150},
  url = {https://link.aps.org/doi/10.1103/PhysRevB.95.235150}
}

@Article{Mizokawa2000,
  title = {Orbital polarons and ferromagnetic insulators in manganites},
  author = {Mizokawa, T. and Khomskii, D. I. and Sawatzky, G. A.},
  journal = {Phys. Rev. B},
  volume = {63},
  issue = {2},
  pages = {024403},
  numpages = {5},
  year = {2000},
  month = {Dec},
  publisher = {American Physical Society},
  doi = {10.1103/PhysRevB.63.024403},
  url = {https://link.aps.org/doi/10.1103/PhysRevB.63.024403}
}

@Article{Torrance1992,
  title = {Systematic study of insulator-metal transitions in perovskites R${\mathrm{NiO}}_{3}$ (R=Pr,Nd,Sm,Eu) due to closing of charge-transfer gap},
  author = {Torrance, J. B. and Lacorre, P. and Nazzal, A. I. and Ansaldo, E. J. and Niedermayer, Ch.},
  journal = {Phys. Rev. B},
  volume = {45},
  issue = {14},
  pages = {8209--8212},
  numpages = {0},
  year = {1992},
  month = {Apr},
  publisher = {American Physical Society},
  doi = {10.1103/PhysRevB.45.8209},
  url = {https://link.aps.org/doi/10.1103/PhysRevB.45.8209}
}

@article{Lee2011,
  title = {Polarity control of carrier injection at ferroelectric/metal interfaces for electrically switchable diode and photovoltaic effects},
  author = {Lee, D. and Baek, S. H. and Kim, T. H. and Yoon, J.-G. and Folkman, C. M. and Eom, C. B. and Noh, T. W.},
  journal = {Phys. Rev. B},
  volume = {84},
  issue = {12},
  pages = {125305},
  numpages = {9},
  year = {2011},
  month = {Sep},
  publisher = {American Physical Society},
  doi = {10.1103/PhysRevB.84.125305},
  url = {https://link.aps.org/doi/10.1103/PhysRevB.84.125305}
}

@article{Kanamori1960,
	title = {Crystal {Distortion} in {Magnetic} {Compounds}},
	volume = {31},
	issn = {0021-8979, 1089-7550},
	url = {http://aip.scitation.org/doi/10.1063/1.1984590},
	doi = {10.1063/1.1984590},
	language = {en},
	number = {5},
	urldate = {2020-01-21},
	journal = {Journal of Applied Physics},
	author = {Kanamori, Junjiro},
	month = may,
	year = {1960},
	note = {Number: 5 Reporter: Journal of Applied Physics},
	pages = {S14--S23}
}

@article{Alonso2001,
  title = {High-temperature structural evolution of $R{\mathrm{NiO}}_{3}$ $(R=\mathrm{H}\mathrm{o},\mathrm{}\mathrm{Y},\mathrm{ }\mathrm{E}\mathrm{r},\mathrm{ }\mathrm{Lu})$ perovskites: Charge disproportionation and electronic localization},
  author = {Alonso, J. A. and Mart\'{\i}nez-Lope, M. J. and Casais, M. T. and Garc\'{\i}a-Mu\~noz, J. L. and Fern\'andez-D\'{\i}az, M. T. and Aranda, M. A. G.},
  journal = {Phys. Rev. B},
  volume = {64},
  issue = {9},
  pages = {094102},
  numpages = {10},
  year = {2001},
  month = {Jul},
  publisher = {American Physical Society},
  doi = {10.1103/PhysRevB.64.094102},
  url = {https://link.aps.org/doi/10.1103/PhysRevB.64.094102}
}

@article{Imada1998,
  title = {Metal-insulator transitions},
  author = {Imada, Masatoshi and Fujimori, Atsushi and Tokura, Yoshinori},
  journal = {Rev. Mod. Phys.},
  volume = {70},
  issue = {4},
  pages = {1039--1263},
  numpages = {0},
  year = {1998},
  month = {Oct},
  publisher = {American Physical Society},
  doi = {10.1103/RevModPhys.70.1039},
  url = {https://link.aps.org/doi/10.1103/RevModPhys.70.1039}
}

@article{Stewart2011,
  title = {Mott Physics near the Insulator-To-Metal Transition in ${\mathrm{NdNiO}}_{3}$},
  author = {Stewart, M. K. and Liu, Jian and Kareev, M. and Chakhalian, J. and Basov, D. N.},
  journal = {Phys. Rev. Lett.},
  volume = {107},
  issue = {17},
  pages = {176401},
  numpages = {5},
  year = {2011},
  month = {Oct},
  publisher = {American Physical Society},
  doi = {10.1103/PhysRevLett.107.176401},
  url = {https://link.aps.org/doi/10.1103/PhysRevLett.107.176401}
}

@article{Medarde1997,
	doi = {10.1088/0953-8984/9/8/003},
	url = {https://doi.org/10.1088/0953-8984/9/8/003},
	year = 1997,
	month = {feb},
	publisher = {{IOP} Publishing},
	volume = {9},
	number = {8},
	pages = {1679--1707},
	author = {Mar{\'{\i}}a Luisa Medarde},
	title = {Structural, magnetic and electronic properties of perovskites (R = rare earth)},
	journal = {Journal of Physics: Condensed Matter},
}

@article{Alonso1999,
  title = {Charge Disproportionation in $\mathit{R}{\mathrm{NiO}}_{3}$ Perovskites: Simultaneous Metal-Insulator and Structural Transition in ${\mathrm{YNiO}}_{3}$},
  author = {Alonso, J. A. and Garc\'{\i}a-Mu\~noz, J. L. and Fern\'andez-D\'{\i}az, M. T. and Aranda, M. A. G. and Mart\'{\i}nez-Lope, M. J. and Casais, M. T.},
  journal = {Phys. Rev. Lett.},
  volume = {82},
  issue = {19},
  pages = {3871--3874},
  numpages = {0},
  year = {1999},
  month = {May},
  publisher = {American Physical Society},
  doi = {10.1103/PhysRevLett.82.3871},
  url = {https://link.aps.org/doi/10.1103/PhysRevLett.82.3871}
}

@article{Mazin2007,
  title = {Charge Ordering as Alternative to Jahn-Teller Distortion},
  author = {Mazin, I. I. and Khomskii, D. I. and Lengsdorf, R. and Alonso, J. A. and Marshall, W. G. and Ibberson, R. M. and Podlesnyak, A. and Mart\'{\i}nez-Lope, M. J. and Abd-Elmeguid, M. M.},
  journal = {Phys. Rev. Lett.},
  volume = {98},
  issue = {17},
  pages = {176406},
  numpages = {4},
  year = {2007},
  month = {Apr},
  publisher = {American Physical Society},
  doi = {10.1103/PhysRevLett.98.176406},
  url = {https://link.aps.org/doi/10.1103/PhysRevLett.98.176406}
}

@article{Scagnoli2005,
  title = {Charge disproportionation and search for orbital ordering in $\mathrm{Nd}\mathrm{Ni}{\mathrm{O}}_{3}$ by use of resonant x-ray diffraction},
  author = {Scagnoli, V. and Staub, U. and Janousch, M. and Mulders, A. M. and Shi, M. and Meijer, G. I. and Rosenkranz, S. and Wilkins, S. B. and Paolasini, L. and Karpinski, J. and Kazakov, S. M. and Lovesey, S. W.},
  journal = {Phys. Rev. B},
  volume = {72},
  issue = {15},
  pages = {155111},
  numpages = {7},
  year = {2005},
  month = {Oct},
  publisher = {American Physical Society},
  doi = {10.1103/PhysRevB.72.155111},
  url = {https://link.aps.org/doi/10.1103/PhysRevB.72.155111}
}

@article{Munoz1994,
  title = {Neutron-diffraction study of the magnetic ordering in the insulating regime of the perovskites R${\mathrm{NiO}}_{3}$ (R=Pr and Nd)},
  author = {Garc\'{\i}a-Mu\~noz, J. L. and Rodr\'{\i}guez-Carvajal, J. and Lacorre, P.},
  journal = {Phys. Rev. B},
  volume = {50},
  issue = {2},
  pages = {978--992},
  numpages = {0},
  year = {1994},
  month = {Jul},
  publisher = {American Physical Society},
  doi = {10.1103/PhysRevB.50.978},
  url = {https://link.aps.org/doi/10.1103/PhysRevB.50.978}
}

@article{Rajpurohit2024_3,
	author = {Rajpurohit, Sangeeta and Vennelakanti, Vyshnavi and Kulik, Heather J.},
	date = {2024/10/03},
	doi = {10.1021/acs.jpca.4c05046},
	isbn = {1089-5639},
	journal = {The Journal of Physical Chemistry A},
	journal1 = {The Journal of Physical Chemistry A},
	journal2 = {J. Phys. Chem. A},
	month = {10},
	publisher = {American Chemical Society},
	title = {Improving Predictions of Spin-Crossover Complex Properties through DFT Calculations with a Local Hybrid Functional},
	type = {doi: 10.1021/acs.jpca.4c05046},
	url = {https://doi.org/10.1021/acs.jpca.4c05046},
	year = {2024},
	year1 = {2024}
}

@article{Jernej2013,
	author = {Georges, Antoine and Medici, Luca de&apos; and Mravlje, Jernej},
	doi = {https://doi.org/10.1146/annurev-conmatphys-020911-125045},
	issn = {1947-5462},
	journal = {Annual Review of Condensed Matter Physics},
	number = {Volume 4, 2013},
	pages = {137-178},
	publisher = {Annual Reviews},
	title = {Strong Correlations from Hund's Coupling},
	type = {Journal Article},
	url = {https://www.annualreviews.org/content/journals/10.1146/annurev-conmatphys-020911-125045},
	volume = {4},
	year = {2013}}

@article{Junjiro1963,
    author = {Kanamori, Junjiro},
    title = "{Electron Correlation and Ferromagnetism of Transition Metals}",
    journal = {Progress of Theoretical Physics},
    volume = {30},
    number = {3},
    pages = {275-289},
    year = {1963},
    month = {09},
    issn = {0033-068X},
    doi = {10.1143/PTP.30.275},
    url = {https://doi.org/10.1143/PTP.30.275},
    eprint = {https://academic.oup.com/ptp/article-pdf/30/3/275/5278869/30-3-275.pdf},
}

@article{Mizokawa2000_2,
  title = {Spin and charge ordering in self-doped Mott insulators},
  author = {Mizokawa, T. and Khomskii, D. I. and Sawatzky, G. A.},
  journal = {Phys. Rev. B},
  volume = {61},
  issue = {17},
  pages = {11263--11266},
  numpages = {0},
  year = {2000},
  month = {May},
  publisher = {American Physical Society},
  doi = {10.1103/PhysRevB.61.11263},
  url = {https://link.aps.org/doi/10.1103/PhysRevB.61.11263}
}

@article{Giovannetti2009,
  title = {Multiferroicity in Rare-Earth Nickelates $R{\mathrm{NiO}}_{3}$},
  author = {Giovannetti, Gianluca and Kumar, Sanjeev and Khomskii, Daniel and Picozzi, Silvia and van den Brink, Jeroen},
  journal = {Phys. Rev. Lett.},
  volume = {103},
  issue = {15},
  pages = {156401},
  numpages = {4},
  year = {2009},
  month = {Oct},
  publisher = {American Physical Society},
  doi = {10.1103/PhysRevLett.103.156401},
  url = {https://link.aps.org/doi/10.1103/PhysRevLett.103.156401}
}

@article{Perdew1996,
    author = {Perdew, John P. and Ernzerhof, Matthias and Burke, Kieron},
    title = {Rationale for mixing exact exchange with density functional approximations},
    journal = {The Journal of Chemical Physics},
    volume = {105},
    number = {22},
    pages = {9982-9985},
    year = {1996},
    month = {12},
    issn = {0021-9606},
    doi = {10.1063/1.472933},
    url = {https://doi.org/10.1063/1.472933},
    eprint = {https://pubs.aip.org/aip/jcp/article-pdf/105/22/9982/19228856/9982\_1\_online.pdf},
}

@article{Eckhoff2020,
  title = {Hybrid density functional theory benchmark study on lithium manganese oxides},
  author = {Eckhoff, Marco and Bl\"ochl, Peter E. and Behler, J\"org},
  journal = {Phys. Rev. B},
  volume = {101},
  issue = {20},
  pages = {205113},
  numpages = {16},
  year = {2020},
  month = {May},
  publisher = {American Physical Society},
  doi = {10.1103/PhysRevB.101.205113},
  url = {https://link.aps.org/doi/10.1103/PhysRevB.101.205113}
}

@article{Subedi2015,
  title = {Low-energy description of the metal-insulator transition in the rare-earth nickelates},
  author = {Subedi, Alaska and Peil, Oleg E. and Georges, Antoine},
  journal = {Phys. Rev. B},
  volume = {91},
  issue = {7},
  pages = {075128},
  numpages = {16},
  year = {2015},
  month = {Feb},
  publisher = {American Physical Society},
  doi = {10.1103/PhysRevB.91.075128},
  url = {https://link.aps.org/doi/10.1103/PhysRevB.91.075128}
}

@article{Ruppen2015,
  title = {Optical spectroscopy and the nature of the insulating state of rare-earth nickelates},
  author = {Ruppen, J. and Teyssier, J. and Peil, O. E. and Catalano, S. and Gibert, M. and Mravlje, J. and Triscone, J.-M. and Georges, A. and van der Marel, D.},
  journal = {Phys. Rev. B},
  volume = {92},
  issue = {15},
  pages = {155145},
  numpages = {11},
  year = {2015},
  month = {Oct},
  publisher = {American Physical Society},
  doi = {10.1103/PhysRevB.92.155145},
  url = {https://link.aps.org/doi/10.1103/PhysRevB.92.155145}
}

@article{Lu2017,
  title = {Origins of bond and spin order in rare-earth nickelate bulk and heterostructures},
  author = {Lu, Yi and Zhong, Zhicheng and Haverkort, Maurits W. and Hansmann, Philipp},
  journal = {Phys. Rev. B},
  volume = {95},
  issue = {19},
  pages = {195117},
  numpages = {5},
  year = {2017},
  month = {May},
  publisher = {American Physical Society},
  doi = {10.1103/PhysRevB.95.195117},
  url = {https://link.aps.org/doi/10.1103/PhysRevB.95.195117}
}

@article{Piel2019,
  title = {Mechanism and control parameters of the coupled structural and metal-insulator transition in nickelates},
  author = {Peil, Oleg E. and Hampel, Alexander and Ederer, Claude and Georges, Antoine},
  journal = {Phys. Rev. B},
  volume = {99},
  issue = {24},
  pages = {245127},
  numpages = {9},
  year = {2019},
  month = {Jun},
  publisher = {American Physical Society},
  doi = {10.1103/PhysRevB.99.245127},
  url = {https://link.aps.org/doi/10.1103/PhysRevB.99.245127}
}

@article{Zavaleta2021,
  title = {Effects of reduced dimensionality, crystal field, electron-lattice coupling, and strain on the ground state of a rare-earth nickelate monolayer},
  author = {Chavez Zavaleta, Rodrigo and Fomichev, Stepan and Khaliullin, Giniyat and Berciu, Mona},
  journal = {Phys. Rev. B},
  volume = {104},
  issue = {20},
  pages = {205111},
  numpages = {14},
  year = {2021},
  month = {Nov},
  publisher = {American Physical Society},
  doi = {10.1103/PhysRevB.104.205111},
  url = {https://link.aps.org/doi/10.1103/PhysRevB.104.205111}
}

@article{Varignon2017,
	author = {Varignon, Julien and Grisolia, Mathieu N. and {\'I}{\~n}iguez, Jorge and Barth{\'e}l{\'e}my, Agn{\`e}s and Bibes, Manuel},
	date = {2017/04/19},
	doi = {10.1038/s41535-017-0024-9},
	id = {Varignon2017},
	isbn = {2397-4648},
	journal = {npj Quantum Materials},
	number = {1},
	pages = {21},
	title = {Complete phase diagram of rare-earth nickelates from first-principles},
	url = {https://doi.org/10.1038/s41535-017-0024-9},
	volume = {2},
	year = {2017}
    }

@article{Hampel2017,
  title = {Interplay between breathing mode distortion and magnetic order in rare-earth nickelates $R{\mathrm{NiO}}_{3}$ within $\mathrm{DFT}+U$},
  author = {Hampel, Alexander and Ederer, Claude},
  journal = {Phys. Rev. B},
  volume = {96},
  issue = {16},
  pages = {165130},
  numpages = {12},
  year = {2017},
  month = {Oct},
  publisher = {American Physical Society},
  doi = {10.1103/PhysRevB.96.165130},
  url = {https://link.aps.org/doi/10.1103/PhysRevB.96.165130}
}

@article{Prosandeev2012,
  title = {Ab initio study of the factors affecting the ground state of rare-earth nickelates},
  author = {Prosandeev, Sergey and Bellaiche, L. and \'I\~niguez, Jorge},
  journal = {Phys. Rev. B},
  volume = {85},
  issue = {21},
  pages = {214431},
  numpages = {7},
  year = {2012},
  month = {Jun},
  publisher = {American Physical Society},
  doi = {10.1103/PhysRevB.85.214431},
  url = {https://link.aps.org/doi/10.1103/PhysRevB.85.214431}
}

@article{Hampel2019,
	author = {Hampel, Alexander and Liu, Peitao and Franchini, Cesare and Ederer, Claude},
	date = {2019/02/06},
	doi = {10.1038/s41535-019-0145-4},
	id = {Hampel2019},
	isbn = {2397-4648},
	journal = {npj Quantum Materials},
	number = {1},
	pages = {5},
	title = {Energetics of the coupled electronic--structural transition in the rare-earth nickelates},
	url = {https://doi.org/10.1038/s41535-019-0145-4},
	volume = {4},
	year = {2019}
}

@article{Seth2017,
  title = {Renormalization of effective interactions in a negative charge transfer insulator},
  author = {Seth, Priyanka and Peil, Oleg E. and Pourovskii, Leonid and Betzinger, Markus and Friedrich, Christoph and Parcollet, Olivier and Biermann, Silke and Aryasetiawan, Ferdi and Georges, Antoine},
  journal = {Phys. Rev. B},
  volume = {96},
  issue = {20},
  pages = {205139},
  numpages = {11},
  year = {2017},
  month = {Nov},
  publisher = {American Physical Society},
  doi = {10.1103/PhysRevB.96.205139},
  url = {https://link.aps.org/doi/10.1103/PhysRevB.96.205139}
}

@article{Yamamoto2002,
	author = {Yamamoto ,Susumu and Fujiwara ,Takeo},
	doi = {10.1143/JPSJ.71.1226},
	eprint = {https://doi.org/10.1143/JPSJ.71.1226},
	journal = {Journal of the Physical Society of Japan},
	number = {5},
	pages = {1226-1229},
	title = {Charge and Spin Order in RNiO3 (R=Nd, Y) by LSDA+U Method},
	url = {https://doi.org/10.1143/JPSJ.71.1226},
	volume = {71},
	year = {2002}
}

@article{Horiba2007,
  title = {Electronic structure of $\mathrm{La}\mathrm{Ni}{\mathrm{O}}_{3\ensuremath{-}x}$: An in situ soft x-ray photoemission and absorption study},
  author = {Horiba, K. and Eguchi, R. and Taguchi, M. and Chainani, A. and Kikkawa, A. and Senba, Y. and Ohashi, H. and Shin, S.},
  journal = {Phys. Rev. B},
  volume = {76},
  issue = {15},
  pages = {155104},
  numpages = {6},
  year = {2007},
  month = {Oct},
  publisher = {American Physical Society},
  doi = {10.1103/PhysRevB.76.155104},
  url = {https://link.aps.org/doi/10.1103/PhysRevB.76.155104}
}

@article{Han2012,
  title = {Spin-moment formation and reduced orbital polarization in LaNiO${}_{3}$/LaAlO${}_{3}$ superlattice: $\mathrm{LDA}+U$ study},
  author = {Han, Myung Joon and van Veenendaal, Michel},
  journal = {Phys. Rev. B},
  volume = {85},
  issue = {19},
  pages = {195102},
  numpages = {5},
  year = {2012},
  month = {May},
  publisher = {American Physical Society},
  doi = {10.1103/PhysRevB.85.195102},
  url = {https://link.aps.org/doi/10.1103/PhysRevB.85.195102}
}

@article{Abbate2002,
  title = {Electronic structure and metal-insulator transition in ${\mathrm{LaNiO}}_{3\ensuremath{-}\ensuremath{\delta}}$},
  author = {Abbate, M. and Zampieri, G. and Prado, F. and Caneiro, A. and Gonzalez-Calbet, J. M. and Vallet-Regi, M.},
  journal = {Phys. Rev. B},
  volume = {65},
  issue = {15},
  pages = {155101},
  numpages = {6},
  year = {2002},
  month = {Mar},
  publisher = {American Physical Society},
  doi = {10.1103/PhysRevB.65.155101},
  url = {https://link.aps.org/doi/10.1103/PhysRevB.65.155101}
}

@article{Lau2013,
  title = {Theory of the Magnetic and Metal-Insulator Transitions in $R{\mathrm{NiO}}_{3}$ Bulk and Layered Structures},
  author = {Lau, Bayo and Millis, Andrew J.},
  journal = {Phys. Rev. Lett.},
  volume = {110},
  issue = {12},
  pages = {126404},
  numpages = {6},
  year = {2013},
  month = {Mar},
  publisher = {American Physical Society},
  doi = {10.1103/PhysRevLett.110.126404},
  url = {https://link.aps.org/doi/10.1103/PhysRevLett.110.126404}
}

@article{Park2012,
  title = {Site-Selective Mott Transition in Rare-Earth-Element Nickelates},
  author = {Park, Hyowon and Millis, Andrew J. and Marianetti, Chris A.},
  journal = {Phys. Rev. Lett.},
  volume = {109},
  issue = {15},
  pages = {156402},
  numpages = {5},
  year = {2012},
  month = {Oct},
  publisher = {American Physical Society},
  doi = {10.1103/PhysRevLett.109.156402},
  url = {https://link.aps.org/doi/10.1103/PhysRevLett.109.156402}
}

@article{GuzmanVerri2019,
	author = {Guzm{\'a}n-Verri, G. G. and Brierley, R. T. and Littlewood, P. B.},
	date = {2019/12/01},
	doi = {10.1038/s41586-019-1824-9},
	id = {Guzm{\'a}n-Verri2019},
	isbn = {1476-4687},
	journal = {Nature},
	number = {7787},
	pages = {429--432},
	title = {Cooperative elastic fluctuations provide tuning of the metal--insulator transition},
	url = {https://doi.org/10.1038/s41586-019-1824-9},
	volume = {576},
	year = {2019}
    }

@article{Piel2014,
  title = {Orbital polarization in strained ${\mathrm{LaNiO}}_{3}$: Structural distortions and correlation effects},
  author = {Peil, Oleg E. and Ferrero, Michel and Georges, Antoine},
  journal = {Phys. Rev. B},
  volume = {90},
  issue = {4},
  pages = {045128},
  numpages = {12},
  year = {2014},
  month = {Jul},
  publisher = {American Physical Society},
  doi = {10.1103/PhysRevB.90.045128},
  url = {https://link.aps.org/doi/10.1103/PhysRevB.90.045128}
}

@article{He2015,
  title = {Strain control of electronic phase in rare-earth nickelates},
  author = {He, Zhuoran and Millis, Andrew J.},
  journal = {Phys. Rev. B},
  volume = {91},
  issue = {19},
  pages = {195138},
  numpages = {9},
  year = {2015},
  month = {May},
  publisher = {American Physical Society},
  doi = {10.1103/PhysRevB.91.195138},
  url = {https://link.aps.org/doi/10.1103/PhysRevB.91.195138}
}

@article{Adamo1999,
    author = {Adamo, Carlo and Barone, Vincenzo},
    title = {Toward reliable density functional methods without adjustable parameters: The PBE0 model},
    journal = {The Journal of Chemical Physics},
    volume = {110},
    number = {13},
    pages = {6158-6170},
    year = {1999},
    month = {04},
    issn = {0021-9606},
    doi = {10.1063/1.478522},
    url = {https://doi.org/10.1063/1.478522},
}

@ARTICLE{Rajpurohit2026,
       author = {{Rajpurohit}, Sangeeta and {Haque}, Sheikh Rubaiat Ul and {Lindenberg}, Aaron M. and {Bl{\"o}chl}, Peter E. and {Ogitsu}, Tadashi},
        title = "{Photoinduced orbital polarization and Jahn-Teller effect in RNiO$_3$}",
      journal = {arXiv e-prints},
         year = 2026,
        month = apr,
          eid = {arXiv:2604.18524},
        pages = {arXiv:2604.18524},
          doi = {10.48550/arXiv.2604.18524},
archivePrefix = {arXiv},
       eprint = {2604.18524},
 primaryClass = {cond-mat.str-el},
       adsurl = {https://ui.adsabs.harvard.edu/abs/2026arXiv260418524R}
}

@article{Dejean2018,
  title = {Ultrafast Modification of Hubbard $U$ in a Strongly Correlated Material: Ab initio High-Harmonic Generation in NiO},
  author = {Tancogne-Dejean, Nicolas and Sentef, Michael A. and Rubio, Angel},
  journal = {Phys. Rev. Lett.},
  volume = {121},
  issue = {9},
  pages = {097402},
  numpages = {6},
  year = {2018},
  month = {Aug},
  publisher = {American Physical Society},
  doi = {10.1103/PhysRevLett.121.097402},
  url = {https://link.aps.org/doi/10.1103/PhysRevLett.121.097402}
}

@article{Baykusheva2022,
  title = {Ultrafast Renormalization of the On-Site Coulomb Repulsion in a Cuprate Superconductor},
  author = {Baykusheva, Denitsa R. and Jang, Hoyoung and Husain, Ali A. and Lee, Sangjun and TenHuisen, Sophia F. R. and Zhou, Preston and Park, Sunwook and Kim, Hoon and Kim, Jin-Kwang and Kim, Hyeong-Do and Kim, Minseok and Park, Sang-Youn and Abbamonte, Peter and Kim, B. J. and Gu, G. D. and Wang, Yao and Mitrano, Matteo},
  journal = {Phys. Rev. X},
  volume = {12},
  issue = {1},
  pages = {011013},
  numpages = {14},
  year = {2022},
  month = {Jan},
  publisher = {American Physical Society},
  doi = {10.1103/PhysRevX.12.011013},
  url = {https://link.aps.org/doi/10.1103/PhysRevX.12.011013}
}

@article{Graanas2022,
  title = {Ultrafast modification of the electronic structure of a correlated insulator},
  author = {Gr\aa{}n\"as, O. and Vaskivskyi, I. and Wang, X. and Thunstr\"om, P. and Ghimire, S. and Knut, R. and S\"oderstr\"om, J. and Kjellsson, L. and Turenne, D. and Engel, R. Y. and Beye, M. and Lu, J. and Higley, D. J. and Reid, A. H. and Schlotter, W. and Coslovich, G. and Hoffmann, M. and Kolesov, G. and Sch\"u\ss{}ler-Langeheine, C. and Styervoyedov, A. and Tancogne-Dejean, N. and Sentef, M. A. and Reis, D. A. and Rubio, A. and Parkin, S. S. P. and Karis, O. and Rubensson, J.-E. and Eriksson, O. and D\"urr, H. A.},
  journal = {Phys. Rev. Res.},
  volume = {4},
  issue = {3},
  pages = {L032030},
  numpages = {7},
  year = {2022},
  month = {Aug},
  publisher = {American Physical Society},
  doi = {10.1103/PhysRevResearch.4.L032030},
  url = {https://link.aps.org/doi/10.1103/PhysRevResearch.4.L032030}
}

\appendix

\section{Hopping matrix element}
The hopping matrix elements along $x$, $y$ and $z$ directions are defined as  
\begin{eqnarray}
T^{x}_{R,R'}&=&
-\frac{1}{4}t_{hop}
\left(\begin{array}{cc}
  3   & -\sqrt{3} \\
  -\sqrt{3}   & 1
\end{array}
\right)
\\
T^{y}_{R,R'}&=&
-\frac{1}{4}t_{hop}
\left(\begin{array}{cc}
  3   & +\sqrt{3} \\
  +\sqrt{3}   & 1
\end{array}
\right)
\\
T^{z}_{R,R'}&=&
t_{hop}
\left(\begin{array}{cc}
  3   & 0 \\
  0  & 3
\end{array}
\right)
\end{eqnarray}

We include three octahedral phonon modes in our model, 
two JT ($Q_{2,R}$ and $Q_{3,R}$) and a
breathing $Q_{1,R}$ \cite{Kanamori1960}.
These phonon modes are defined as 
\begin{eqnarray}
Q_{1,R} &=& \frac{1}{\sqrt{3}}(d_{x,R}+d_{y,R}+d_{z,R}-3\bar{d}) \\
Q_{2,R} &=& \frac{1}{\sqrt{2}}(d_{x,R}-d_{y,R}) \\
Q_{3,R} &=& \frac{1}{\sqrt{6}}(2d_{z,R}-d_{x,R}-d_{y,R}),
\label{eq:tbm_Jahn_Teller_modes}
\end{eqnarray}
where $d_{x,R}$, $d_{y,R}$ and $d_{z,R}$ denote the
O-O distances along $x$, $y$, and $z$ directions, 
respectively, surrounding the Ni site $R$. 
$\bar{d}=3.97$ $\AA$ is the equilibrium O-O distance.

Table \ref{tab:simulation_setup} summarizes the relevant parameters that we used for the simulations.
\begin{table}[!htb]
\centering 
\label{tab:t2}
\begin{tabular}{|l|l|}
\hline\hline
k-grid    & $4{\times}4{\times}4$ \\
supercell  & $N_x{\times}N_y{\times}N_z{=}4{\times}4{\times}4$\\ 
Ni sites per unit cell  &$N_{Ni}=64$\\
O sites per unit cell  &$N_{O}=192$\\
Ni-Ni spacing & $d_{Ni-Ni}=3.97$~\AA\\ 
\hline
\hline
\end{tabular}
\caption{Simulation set up used for the results presented in the main text.}
\label{tab:simulation_setup}
\end{table}

\section{k-point  convergence}
To test k-point convergence, we examine the total energy of the CO-OO state for different k-grid sizes.
Table \ref{tab:convergence_table} shows the total energy for different k-point grids, demonstrating k-point convergence.
\begin{table}[!htb]
\centering
\label{tab:t3}
\begin{tabular}{|l|l|}
\hline\hline
k-grid   & Total energy (eV) \\
\hline
$1{\times}1{\times}1$ & $-0.1106$ eV/Ni \\
$2{\times}2{\times}2$ & $-0.1083$ eV/Ni \\
$3{\times}3{\times}3$ & $-0.1083$ eV/Ni \\
$4{\times}4{\times}4$ & $-0.1070$ eV/Ni \\
\hline
\hline
\end{tabular}
\caption{k-point convergence test: total energy as a function of k-grid density.}
\label{tab:convergence_table}
\end{table}

\end{document}